\newif\ifNotes\Notesfalse
\newif\ifAnon\Anonfalse
\newif\ifDraft\Draftfalse
\newif\ifOld\Oldtrue
\newif\ifFull\Fulltrue

\documentclass[letterpaper,twocolumn,10pt]{article}
\usepackage[hyphens]{url}
\usepackage[table]{xcolor}
\usepackage{usenix2019_v3,epsfig,endnotes}
\usepackage{color}
\usepackage[numbers,sort]{natbib}
\usepackage{appendix}

\usepackage{booktabs} % For formal tables
\usepackage{xspace}
\usepackage{textcomp}
\usepackage[hyphens]{url}
\usepackage{hyperref}
\usepackage[flushleft]{threeparttable}
\usepackage{pifont}
\usepackage{courier}
\usepackage{graphicx}
\usepackage{array}
\usepackage{listings}
\usepackage{amsmath}
\usepackage[geometry]{ifsym}

\usepackage{enumitem}
\usepackage{tikz-uml}
\usepackage{caption}
\usepackage{subcaption}
\usepackage[symbol]{footmisc}

\setitemize{itemsep=1ex,topsep=1ex,parsep=0pt,partopsep=0pt}

\let\origthelstnumber\thelstnumber
\makeatletter
\newcommand*\Suppressnumber{%
  \lst@AddToHook{OnNewLine}{%
    \let\thelstnumber\relax%
     \advance\c@lstnumber-\@ne\relax%
    }%
}

\newcommand*\Reactivatenumber[1]{%
  \setcounter{lstnumber}{\numexpr#1-1\relax}
  \lst@AddToHook{OnNewLine}{%
   \let\thelstnumber\origthelstnumber%
   \refstepcounter{lstnumber}
  }%
}

\makeatother

\ifDraft
  \usepackage{draftwatermark}
  \definecolor{watermarkcolor}{rgb}{0.8,0.8,1}
  \SetWatermarkText{DRAFT}
  \SetWatermarkColor{watermarkcolor}
\fi

\definecolor{linkcolor}{rgb}{0.65,0,0}
\definecolor{citecolor}{rgb}{0,0.4,0}
\definecolor{urlcolor}{rgb}{0,0,0.65}
\definecolor{TolDarkGreen}{HTML}{117733}
\hypersetup{hyperindex=true,pdfpagemode=UseNone,pdfstartview=FitH,colorlinks=true, linkcolor=linkcolor, urlcolor=urlcolor, citecolor=citecolor}

\newcommand{\twocol}[1]{\multicolumn{2}{l}{#1}}

\usepackage{color}

\usepackage[english]{babel}
\usepackage{blindtext}
\usepackage{tikz}

\newcommand{\swallow}[1]{}
\ifNotes
  \newcommand{\colorcomment}[2]{\leavevmode\unskip\space{\color{#1}#2}\xspace}
\else
  \newcommand{\colorcomment}[2]{\leavevmode\unskip\relax}
\fi

\newcommand{\noArxiv}{\relax}

\newcommand{\taggedcolorcomment}[3]{\noArxiv\colorcomment{#1}{[\textbf{#2}: #3]}}

\newcommand{\yval}[1]{\taggedcolorcomment{purple}{yval}{#1}}

\newcommand{\tpm}{$\pm$}

\newcommand{\xmark}{\leavevmode{\color{red}\ding{55}}\xspace}%
\newcommand{\cmark}{\leavevmode{\color{TolDarkGreen}\ding{51}}\xspace}%

\ifOld
  \newcommand{\old}[1]{\noArxiv\leavevmode\unskip\relax}
  
\else
  \newcommand{\old}[1]{\noArxiv\leavevmode\unskip\relax}
  
\fi

\definecolor{lightgray}{rgb}{0.95, 0.95, 0.95}
\definecolor{darkgray}{rgb}{0.4, 0.4, 0.4}
\definecolor{editorGray}{rgb}{0.95, 0.95, 0.95}
\definecolor{editorOcher}{rgb}{1, 0.5, 0} % #FF7F00 -> rgb(239, 169, 0)
\definecolor{editorGreen}{rgb}{0, 0.5, 0} % #007C00 -> rgb(0, 124, 0)
\definecolor{orange}{rgb}{1,0.45,0.13}		
\definecolor{olive}{rgb}{0.17,0.59,0.20}
\definecolor{brown}{rgb}{0.69,0.31,0.31}
\definecolor{purple}{rgb}{0.38,0.18,0.81}
\definecolor{lightblue}{rgb}{0.1,0.57,0.7}
\definecolor{lightred}{rgb}{1,0.4,0.5}

\lstdefinelanguage{JavaScript}{
  morekeywords={typeof, new, true, false, catch, function, return, null, catch, switch, var, if, in, while, do, else, case, break, const},
  morecomment=[s]{/*}{*/},
  morecomment=[l]//,
  morestring=[b]",
  morestring=[b]',
  morestring=[b]/,
  stringstyle=\color{olive}
}

\lstdefinelanguage{HTML5}{
  language=html,
  sensitive=true,	
  alsoletter={<>=-},	
  morecomment=[s]{<!-}{-->},
  tag=[s],
  otherkeywords={
	<!DOCTYPE,
  </html, <html, <head, <title, </title, <style, </style, <link, </head, <meta, />,
	</body, <body,
	</div, <div, </div>, 
	</p, <p, </p>,
	</script>, <script>,
  <canvas, /canvas>, <svg, <rect, <animateTransform, </rect>, </svg>, <video, <source, <iframe, ></iframe>, </video>, <image, </image>, <header, </header, <article, </article
  },
  ndkeywords={
  =,
  charset=, src=, id=, width=, height=, style=, type=, rel=, href=,
  fill=, attributeName=, begin=, dur=, from=, to=, poster=, controls=, x=, y=, repeatCount=, xlink:href=,
  margin:, padding:, background-image:, border:, top:, left:, position:, width:, height:, margin-top:, margin-bottom:, font-size:, line-height:,
  transform:, -moz-transform:, -webkit-transform:,
  animation:, -webkit-animation:,
  transition:,  transition-duration:, transition-property:, transition-timing-function:,
  }
}

\lstdefinestyle{web} {
  basicstyle={\footnotesize\ttfamily},   
  frame=single,
  xleftmargin={0.75cm},
  numbers=left,
  stepnumber=1,
  firstnumber=1,
  numberfirstline=true,	
  identifierstyle=\color{black},
  keywordstyle=\color{blue}\bfseries,
  ndkeywordstyle=\color{editorGreen}\bfseries,
  stringstyle=\color{editorOcher}\ttfamily,
  commentstyle=\color{brown}\ttfamily,
  language=HTML5,
  alsolanguage=JavaScript,
  alsodigit={.:;},	
  tabsize=2,
  showtabs=false,
  showspaces=false,
  showstringspaces=false,
  extendedchars=true,
  breaklines=true,
  literate=%
  {Ö}{{\"O}}1
  {Ä}{{\"A}}1
  {Ü}{{\"U}}1
  {ß}{{\ss}}1
  {ü}{{\"u}}1
  {ä}{{\"a}}1
  {ö}{{\"o}}1
}

\newcommand{\parhead}[1]{\vspace{2pt plus 1pt minus 1pt}\par\noindent\textbf{#1}\hspace{.75em plus .5em minus .5em}}

\newcommand{\pp}{Prime+\allowbreak Probe\xspace}

\usepackage[nameinlink,capitalize]{cleveref}
\crefname{figure}{Figure}{Figures}
\crefname{equation}{Equation}{Equations}
\crefformat{equation}{#2{}Equation~#1{}#3}
\crefrangeformat{line}{Lines~#3#1#4--#5#2#6}

\newcommand{\crz}{Chrome Zero\xspace}
\newcommand{\js}{JavaScript\xspace}

\newcommand{\policy}[1]{\textsf{#1}\xspace}
\newcommand{\poloff}{\policy{None}}
\newcommand{\pollow}{\policy{Low}}
\newcommand{\polmed}{\policy{Medium}}
\newcommand{\polhigh}{\policy{High}}
\newcommand{\polpara}{\policy{Paranoid}}

\newcommand{\moffset}{\mathit{offset}}
\newcommand{\msize}{\text{size}}
\newcommand{\mnewsize}{\text{new\_size}}

  \ifAnon
    \newcommand{\authorlist}[1]{\relax}
  \else
    \newcommand{\authorlist}[1]{\author{#1}}
  \fi
  \newcommand{\nextauthor}{\and}
  \newcommand{\myAuthor}[3]{
    {\rm #1} \\
    #2\\
    #3
  }

\begin{document}

\title{Prime+Probe 1, \js 0: Overcoming Browser-based Side-Channel Defenses \\\large\normalfont{(Extended Version)}}

\newlength{\bl}
\setlength{\bl}{2in}
\authorlist{
  \myAuthor{Anatoly Shusterman}{\makebox[\bl]{Ben-Gurion Univ.\ of the Negev}}{shustera@post.bgu.ac.il}\\\\
  \myAuthor{Daniel Genkin}{University of Michigan}{genkin@umich.edu}
  \nextauthor
  \myAuthor{Ayush Agarwal}{University of Michigan}{ayushagr@umich.edu}\\\\
  \myAuthor{Yossi Oren}{\makebox[\bl]{Ben-Gurion Univ.\ of the Negev}}{yos@bgu.ac.il}
  \nextauthor
  \myAuthor{Sioli O'Connell}{University of Adelaide}{\makebox[\bl]{sioli.oconnell@adelaide.edu.au}}\\\\
  \myAuthor{Yuval Yarom}{\makebox[\bl]{University of Adelaide and Data61}}{yval@cs.adelaide.edu.au}
}

  \maketitle

\begin{abstract}
  The ``eternal war in cache'' has reached browsers, with multiple cache-based
  side-channel attacks and countermeasures being suggested.
  A common approach for countermeasures is to disable or restrict \js
  features deemed essential for carrying out attacks.

  To assess the effectiveness of this approach, in this work we seek to
  identify those \js features which are \emph{essential} for carrying
  out a cache-based attack.
  We develop a sequence of attacks with progressively decreasing
  dependency on \js features, culminating in the
  first browser-based side-channel attack which is
  constructed entirely from Cascading Style Sheets (CSS) and HTML,
  and works even when script execution is completely blocked. 
  We then show that 
  avoiding \js features makes our techniques architecturally agnostic, 
  resulting in microarchitectural website fingerprinting attacks that work across hardware platforms
  including Intel Core, AMD Ryzen, Samsung Exynos, and Apple M1 architectures.

  As a final contribution, we evaluate our techniques in hardened browser environments including the Tor browser, DeterFox (Cao el al., CCS 2017), and \crz (Schwartz et al., NDSS 2018).
  We confirm that none of these approaches completely defend against our attacks.
  We further argue that the protections of \crz need to be more comprehensively applied, and that the performance and user experience of \crz will be severely degraded if this approach is taken.
\end{abstract}

\section{Introduction}\label{s:introduction}

The rise in the importance of the web browser in modern society has been
accompanied by an increase in the sensitivity of the information the browser processes.
Consequently, browsers have become targets of attacks aiming to extract or gain
control of users' private information. 
Beyond attacks that target software vulnerabilities and attacks that attempt to profile the device or the user via sensor APIs, browsers have also been used as a platform for mounting microarchitectural side-channel attacks~\cite{GeYCH18}, which recover secrets by measuring the contention on microarchitectural CPU components.

While traditionally such attacks
were implemented using native code~\cite{YanFT20,BrasserMDKCS17,
	OsvikST06,Percival05,YaromF14,LiuYGHL15,GrussSM15}, recent works have
demonstrated that \js code in browsers can also be used to launch such
attacks~\cite{OrenKSK15,GenkinPTY18,GrussMM15,ShustermanKHMMOY19}.
In an attempt to mitigate  \js-based side-channel leakage, browser vendors have
mainly focused on restricting the ability of an attacker to precisely measure
time~\cite{low_resolution_1,low_resolution_2,low_resolution_3}. 

\begin{table*}[htb]
  \resizebox{\linewidth}{!}{
    \begin{tabular}{@{}lllll@{}}
      \toprule
      Countermeasure                & Chrome Zero  & Can Be    & Technique                                                        & External                   \\
                                    & Policy Level & Bypassed? &                                                                  & Requirements               \\
      \midrule
      None                          & \poloff      & \cmark    & Cache Contention \cite{OrenKSK15,GenkinPTY18,ShustermanKHMMOY19} & None                       \\
      Reduced timer resolution      & \polmed      & \cmark    & Sweep Counting \cite{ShustermanKHMMOY19}                         & None                       \\
      No timers, no threads         & \polpara     & \cmark    & DNS Racing                                                       & Non-Cooperating DNS server \\
      No timers, threads, or arrays & ---          & \cmark    & String and Sock                                                  &
      Cooperating WebSockets server                                                                                                                            \\
      \js completely blocked        & ---          & \cmark    & CSS \pp                                                          & Cooperating
      DNS server                                                                                                                                               \\
      \bottomrule
    \end{tabular}
  }
  \caption{Summary of results: \pp Attacks can be Mounted Despite Strict Countermeasures}
\label{t:SummaryOfResults}
\end{table*}

Side-channel attackers,
in turn, attempt to get around these restrictions by creating makeshift timers
with varying accuracies through the exploitation of other browser APIs, such as
message passing or
multithreading~\cite{SchwarzMGM17,kohlbrenner2016trusted,van2015clock}.
More recently, \citet{SchwarzLG18} presented \crz, a Chrome extension
that 
protects against \js-based side-channels by blocking or restricting 
 parts of the \js API commonly used by side channel attackers, based on a
user-selected protection policy. 
Going even further, DeterFox~\cite{CaoCLW17} aims to eliminate side-channel attacks
by ensuring completely deterministic \js execution, 
and NoScript~\cite{noscript} prevents \js-based attacks by completely disabling \js.

A common trend in these approaches is that they are symptomatic
and fail to address the root cause of the leakage, namely, the sharing of microarchitectural
resources. Instead, most approaches attempt to prevent leakage by modifying browser behavior, striking different balances between security and usability.
Thus, we ask the following question.

\medskip
\emph{
  What are the minimal features required for mounting microarchitectural side-channel attacks in browsers? 
  Can attacks be mounted in highly-restricted browser environments, despite security-orientated API refinements?
}
\medskip

Besides being influenced by  defenses, microarchitectural attacks are also affected by
an increased hardware diversification in consumer  devices. 
While the market for high-end processors used to be dominated by Intel,
the past few years have seen an increase in popularity of other alternatives, such as AMD's Zen architecture, Samsung's Exynos, and the recently launched Apple M1 cores.

Most microarchitectural attack techniques, however,  are inherently dependent on the specifics of the underlying CPU hardware, and are typically demonstrated on Intel-based machines. While microarchitectural attacks on non-Intel hardware do exist~\cite{DBLP:conf/uss/LippGSMM16, DBLP:conf/ccs/ZhangXZ16}, these are also far from universal, and are also highly tailored to their respective hardware platforms. Thus, given the ever increasing microarchitectural diversification, we ask the following secondary question.

\medskip
\emph{
	Can microarchitectural side-channel attacks become architecturally-agnostic? In particular, are there universal side channel attacks that can be mounted effectively across diverse architectures, without requiring hardware-dependent modifications?
}

\subsection{Our Contribution}
Tackling the first set of questions, in this paper we show that side channel attacks can be mounted in highly restricted browser environments, despite side-channel hardening of large portions of \js's timing and memory APIs. Moreover, we show that even if \js is  \emph{completely disabled},
side-channel attacks are still possible, albeit with a lower accuracy. We thus
argue that completely preventing side channels in today's
browsers is nearly impossible, with leakage prevention requiring more drastic
design changes. 

Next, tackling the second set of questions, we introduce architecturally-agnostic side channel  techniques, that can operate on highly diverse architectures from different vendors. Empirically evaluating this claim, we show side channel leakage from browser environments running on AMD, Apple, ARM and Intel architectures with virtually no hardware-specific modifications. Notably, to the best of our knowledge, this is the first side-channel attack on Apple's M1 CPU.

\parhead{Reducing Side Channel Requirements.}
We focus our investigation on website fingerprinting attacks~\cite{Hintz02}.
In these attacks, an adversary attempts to breach the privacy of the victim
by finding out the websites that the victim visits.
While initially these attacks relied on network traffic analysis, several past works demonstrated
that an attacker-controlled website running on the victim machine can determine the identity of
other websites the victim visits~\cite{OrenKSK15, NotSoIncognito, KimL016, VilaK17, MatyuninWASK19}.

To identify the set of \js features required for cache attacks,
we build on the work of \cite{ShustermanKHMMOY19}.
We start from their website fingerprinting attacks and design a
sequence of new attacks, each requiring progressively less \js features.
Our process of progressively reducing \js features culminates in CSS \pp, 
which is a microarchitectural attack implemented solely in CSS and HTML,
yet is capable of achieving a high accuracy even when \js is completely disabled.
To the best of our knowledge, this is the first microarchitectural attack
with such minimal requirements.

\parhead{Architecturally-Agnostic Side Channel Attacks.}
Next, we tackle the challenge of mounting side channel attacks across a large variety of computing architectures. We show that the reduced requirements of our techniques essentially make them architecturally-agnostic, allowing them to run on highly diverse architectures with little adaptation. Empirically demonstrating this, we evaluate our attacks on AMD's Ryzen,  Samsung's Exynos and Apple's M1 architectures.  Ironically, we show that our attacks are sometimes more effective on these novel  
CPUs by Apple and Samsung compared to their well-explored Intel counterparts, presumably due to their 
simpler cache replacement policies.

\parhead{Evaluating Existing Side Channel Protections.}
Having reduced the requirements for mounting side channel attacks in browser contexts, we tackle the question of evaluating the security guarantees offered by existing API hardening techniques.
To that aim, we deploy \crz~\cite{SchwarzLG18} and measure
the attack accuracy in the presence of multiple security policies.
We show that while disabling or modifying \js features does attenuate
published attacks, it does little to block attacks that do not
require the disabled features.

As a secondary contribution, we find that there are several gaps in the protection offered by \crz,
and
that fixing those adversely affects \crz's usability and performance.
This raises questions
on the applicability of the approach suggested in \cite{SchwarzLG18} for
protecting browsers.

\parhead{Attacking Hardened Browsers.} Having shown the efficacy of our techniques in both Chrome and \crz environments, we also evaluate our attacks
on several popular security-oriented browsers, such as the Tor Browser~\cite{TorBrowser} and DeterFox~\cite{CaoCLW17}. Here, we show that attacks are still possible, albeit at lower accuracy levels.

\parhead{Summary of Contribution.} In summary, in this paper we make the
following contributions:
\begin{itemize}[leftmargin=*,nolistsep]
  \item We design three cache-based side-channel attacks on browsers, under
        progressively more restrictive assumptions. In particular, we demonstrate
        the first side-channel attack in a browser that does not rely on \js or any
        other mobile code (\cref{s:cache-attacks-without-timers}).
        
        \item We empirically demonstrate architecturally-agnostic side channel attacks, showing the first techniques that can handle diverse architectures with little adaptation (\cref     
        {s:results}).

  \item We re-evaluate \crz's \js API-hardening approach, demonstrating
        significant limitations that affect security, usability, and performance
        (\cref{sec:attacking-crz}).
        
        \item We evaluate our attacks in multiple scenarios, including in the
        restrictive environments of the Tor Browser and DeterFox
        (\cref{sec:harden}).
\end{itemize}

\subsection{Responsible Disclosure} Following the practice of responsible disclosure, we have shared a draft of this paper with the product security teams of Intel, AMD, Apple, Chrome and Mozilla prior to publication.

\section{Background}
\subsection{Microarchitectural Attacks}\label{s:bgattacks}
To improve performance, modern processors typically exploit the {locality} principle, which notes the  tendency of software to reuse the same set of resources within a short period of time. Utilizing this, the processor maintains {state} that describes past program behavior, and uses it for predicting future behavior.

\parhead{Microarchitectural Side Channels.}
The shared use of a processor, therefore, creates the opportunity for
information leakage between programs or security domains~\cite{GeYCH18}. Leakage
could be via shared state~\cite{YaromF14,LeeSGKKP17,AciicmezKS07,GullaschBK11}
or via contention on either the limited state storage
space~\cite{OsvikST06,Percival05,LiuYGHL15,GrasRBG18} or the bandwidth of
microarchitectural components~\cite{YaromGH16,AciicmezS07,AldayaBHGT19}. Exploiting this leakage,  multiple {side-channel attacks} have been presented, extracting cryptographic keys~\cite{YaromF14,OsvikST06, Percival05, LiuYGHL15, YaromGH16, AciicmezS07, AldayaBHGT19, GenkinPSYZ20, RonenGGSWY19, AldayaGTB19, GullaschBK11}, monitoring user behavior~\cite{GrussSM15,OrenKSK15,ShustermanKHMMOY19,GulmezogluZTI0S19,RistenpartTSS09}, and extracting other secret
information~\cite{YanFT20,BrasserMDKCS17,HundWH13}.

Side-channel attacks were shown to allow leaking between
processes~\cite{OsvikST06,Percival05,LiuYGHL15,YaromF14, GullaschBK11}, web
browser tabs~\cite{OrenKSK15,ShustermanKHMMOY19,GenkinPTY18}, virtual
machines~\cite{LiuYGHL15,YaromF14,ZhangJRR12,InciGIES16}, and other security
boundaries~\cite{HundWH13,LeeSGKKP17,BrasserMDKCS17,DallMEGHMY18}.
In this work we are mostly interested in the two attack techniques that target
the limited storage in caching elements, mainly data caches.

\parhead{\pp.}
The {\pp}
attack~\cite{OsvikST06,Percival05,LiuYGHL15} exploits the set-associative
structure in modern caches. The attacker first creates an \emph{eviction set},
which consists of multiple memory locations that map to a single cache set. The
attacker then \emph{primes} the cache by accessing the locations in the eviction
set, filling the cache set with their contents. Finally, the
attacker \emph{probes} the cache by measuring the access time to the eviction
set. A long access time indicates that the victim has accessed memory locations
that map to the same cache set, evicting part of the attacker's data, and
therefore teaches the attacker about the victim's activity.

\parhead{Cache Occupancy.}
In the {cache occupancy} attack~\cite{MauriceNHF15,ShustermanKHMMOY19}, the
attacker repeatedly accesses a cache-sized buffer while measuring the access
time. Because the buffer consumes the entire cache, the access time to the
buffer correlates with the victim's memory activity. The cache occupancy attack
is simpler than \pp, and provides the attacker with less detailed spatial and
temporal information. It is also less sensitive to the clock
resolution~\cite{ShustermanKHMMOY19}. \emph{Sweep
  counting} is a variant of the cache occupancy attack,
in which the adversary counts the number of times that the buffer can be
accessed between two clock ticks. The main advantage of this technique is that
it can work with even lower-resolution clocks.

\subsection{Defenses}
The root cause of microarchitectural side-channels is the sharing of
microarchitectural components across code executing in different protection domains. Hence, partitioning the state, either spatially or
temporally, can be effective in preventing attacks~\cite{GeYCH19}. Partitioning
can be done in hardware~\cite{WangL07,DomnitserJLAP12} or by the operating
system~\cite{LiuGYMRHL16,ShiSCZ11,LiedtkeHH97,KimPM12}.

Fuzzing or reducing the resolution of the clock are often suggested as a
countermeasure~\cite{vattikonda2011eliminating,low_resolution_2,low_resolution_3,Hu91}.
However, these approaches are less effective against the cache occupancy attack,
as it does not require high-resolution timers. Furthermore, these approaches
only introduce uncorrelated noise to the channel and do not prevent
leakage~\cite{CockGMH14}.

Randomizing the cache architecture is another commonly suggested
countermeasure~\cite{WangL07,WernerUG0GM19,Qureshi18}. These often aim to
prevent eviction set creation. However, they are less effective against the
cache occupancy attack, both because the attack does not require eviction sets
and because these techniques do not change the overall cache pressure.

\subsection{The \js Types and Inheritance}\label{s:bgjstype}

\parhead{\js Typing.}
\js is an object oriented language where every value is an object, excluding several basic primitive types.  For object typing, \js mostly uses ``duck typing'',
where an object is considered to have a required type as soon as it has the
expected methods or properties. \js deviates from this model for some built-in
types, such as \texttt{TypedArrays}, which are arrays of primitive types. While
\js code mostly uses these built-in types equivalently to objects, the \js
engine itself provides certain APIs that match the arguments against the
required built-in types, raising exceptions if they mismatch.

\parhead{\js Inheritance.}
\js uses a {prototypal inheritance} model, where each object can have a
single {prototype object}. When searching for a property of an object, \js
first checks the object itself. If the property is not found on in the object,
\js proceeds to check its prototype, until it either finds the property or
reaches an object that has no prototype. The list of prototypes used in this
search is called the object's {prototype chain}. Finally, when \js modifies an object property, the prototype chain is not consulted.
Instead, \js sets the property on the object itself, creating it if it does not
already exist.

\subsection{Virtual Machine Layering}
Virtual machine layering~\cite{LavoieDF14} is a low overhead technique for implementing
function call interception. To intercept calls to a
particular function, the function is overwritten with a new function, in effect
intercepting calls to the original function.

To partially override the behavior of the original function, a reference to the
original function is stored, and the desired behavior is delegated to it if needed.
To prevent external access to the original intercepted function, a \js closure
is used to store this reference. \js closures create new variable scopes,
preventing code outside the closure from accessing references stored within the
closure.

Virtual machine layering offers a significant advantage over other techniques
for guaranteeing that all calls to a given \js function are intercepted. This is
because virtual machine layering changes the definition of the function
directly, automatically supporting the interception of function calls from code
generated at runtime.

\section{Overcoming Browser-based Defenses}\label{s:cache-attacks-without-timers}
In this section we present  several novel  browser-based side-channel techniques that are effective against increasing levels of browser defenses. More specifically, we present a series of attacks that progressively require less \js
features, culminating in CSS \pp -- an attack that does not use \js at all and can work
when \js is completely disabled.
To the best of our knowledge, this is the first
side-channel attack implemented solely with HTML and CSS, without the need of \js.

We evaluate the effectiveness of our techniques via \emph{website fingerprinting} attacks in the Chrome browser, which aim to recover pages currently open on the target's machine. Beyond demonstrating accurate fingerprinting levels against the Chrome browser, we show that our attacks are highly portable, and are effective across several different micro-architectures: Intel x86, AMD Ryzen , Samsung Exynos 2100 (ARM), and finally the Apple M1.

\subsection{Methodology and Experimental Setup}\label{sec:setup}
We follow the methodology of \citet{ShustermanKHMMOY19}, where we collect
\emph{memorygrams}, or traces of cache use over the web site load time. We use
these traces to train a deep neural network model, which is then used to
identify web sites based on the corresponding memorygrams. Similarly to
\cite{ShustermanKHMMOY19}, we measure cache activity using both the cache occupancy and 
sweep counting methods (described below). Both of these methods
measures the overall level of cache contention, obviating the need to construct eviction
sets. Finally, we adapt both techniques to progressively more restrictive
environments.
The specific assumptions on attackers' capabilities appear in the
respective sections (\crefrange{sec:dnsracing}{sec:csspp}).

\parhead{The Cache Occupancy Channel.}
To measure the web page's cache activity, we follow past works~\cite{MauriceNHF15,ShustermanKHMMOY19} 
and use the cache occupancy channel.
Specifically, we allocate an LLC-sized
buffer and measure the time to access the entire buffer.
The  victim's access to memory evicts the contents of our buffer
from the cache, introducing delays for our access.
Thus, the time to access our buffer is roughly proportional to
the number of cache lines that the victim uses.

Compared with the \pp attack, the cache occupancy channel does not
provide any spatial information, meaning that the attacker
does not learn any information about the addresses accessed by the victim.
Thus, it is less appropriate for detailed cryptanalytic attacks which need to track the victim at the 
resolution of a single cache set.
However,the cache occupancy attack is simpler than \pp and in particular avoids the need to construct eviction sets. 
It also requires less accurate temporal information, on the order of milliseconds instead of nanoseconds. 
Thus, cache occupancy attacks are better suited to restricted environments, such as those considered in this section.

\parhead{Sweep Counting.}
Sweep counting~\cite{ShustermanKHMMOY19} is a variant of the basic cache occupancy attack,
with reduced temporal resolution. Here, rather then timing the traversal of a cache-sized buffer, the attacker counts the number of sweeps across the buffer than fit within a time unit. While providing even less accuracy than cache occupancy, sweep counting remains effective when used with low-resolution timing sources (e.g., hundreds of milliseconds). Just like the cache occupancy attack, sweep counting does not provide any spatial resolution.

\parhead{Closed World Evaluation.}
Using the channels we describe above, we collect memorygrams of visits to the Alexa Top 100 websites.
We visit each site 100 times, each time collecting a memorygram that spans 30 seconds.
We then evaluate the accuracy of our techniques  in the \emph{closed-world model}, 
where an adversary knows the list of 100 websites and attempts to guess
which one is visited. Here, the base accuracy rate of a random guess
is 1\%, with any higher accuracy indicating the presence of side-channel leakage in
the collected traces.

\parhead{Evaluated Architectures.}
We demonstrate in the attacks described in this section on several different architectures made by multiple hardware vendors. For Intel, we use several machines featuring an 
Intel Core i5-3470 CPU that has a 6\,MiB last-level cache and 20\,GiB memory. The
machines are running Windows 10 with Chrome version~78, and are connected via 
Ethernet to a university network. Next, for AMD, we used six machines equipped with an 
AMD Ryzen 9 3900X 12-Core Processor, which has a 4x16\,MiB last-level cache and 64\,GiB memory. 
These machines were running Ubuntu 20.04 server with Chrome version~88.0, and were connected via Ethernet to a cloud provider network.
For our ARM evaluation we used five Samsung Galaxy S21 5G mobile phones (SM-G991B), featuring an ARM-based Exynos 2100 CPU with an 8\,MiB last-level cache and 8\,GiB memory. 
These phones were running Android 11 with Chrome 88 and were connected via Wi-Fi to a University network.
Finally, for our evaluation on Apple, we used four Apple Mac Mini machines equipped with an Apple M1 CPU with a 12\,MiB 
last-level cache for performance cores and 4\,MiB for efficiency cores. 
The machines were equipped with 16 \,GiB memory and were running MacOS Big Sur version 11.1 together with Chrome 88.0 for arm64. 
These machines were connected via Ethernet to a University network.

\parhead{Machine Learning Methodology.}
As a classifier we use a deep neural network model, with 10-fold cross
validation. See \cref{app:model} for details. 
 Following previous
works~\cite{Caliskan-IslamHLNVYG15,NarayananPGBSSS12}, we report both the most
likely prediction of the classifier and the top 5 predictions, noting that the
base accuracy for the top 5 results is 5\% for the closed-world scenarios, and 34\% for the open world.
The collected data volume of all the
experiments is 27\,GiB consisting of 40 datasets, where each dataset takes
about one week to collect, and each classifier takes on average 30 minutes to train
on a cluster of Nvidia GTX1080 and GTX2080 GPUs.
\yval{Add location for data}
%We plan to make all of our
%datasets publicly available, together with their associated trained machine
%learning classifiers, to assist in the verification and reproduction of our
%results.

\subsection{DNS Racing}\label{sec:dnsracing}
For our first attack, {DNS Racing}, we assume a hypothetical \js engine that
does not provide any timer, neither through an explicit interface nor via
repurposing \js features such as
multithreading~\cite{SchwarzMGM17,kohlbrenner2016trusted}.

\begin{figure}[htb]
  \includegraphics[trim={0cm 0cm 0cm 0cm}, width=\linewidth]{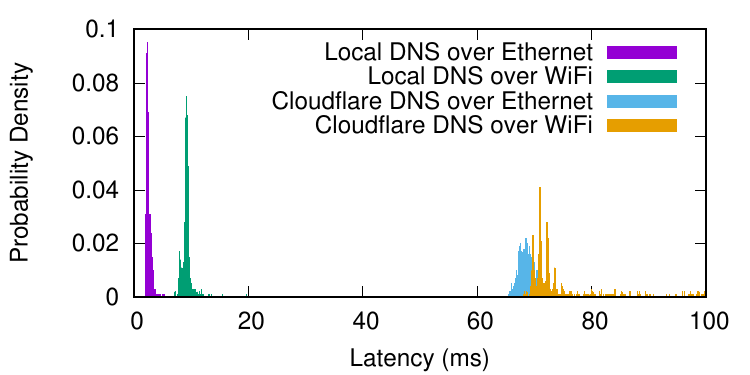}
  \caption{Measured response latencies when loading an image from a non-existent
    domain (local server).
    \label{f:nxdomain-histogram}}
\end{figure}

\parhead{DNS-based Time Measurement.}
\citet{OgenZSO18} observe that browsers behave very predictably when attempting to load a resource from a 
non-existent domain, waiting for exactly one network round-trip before returning an error. 
Thus, it is possible to create  an external timer by setting the \texttt{onerror} handler on an image whose URL 
points to a non-existent domain. We evaluate this timer with a
local DNS server and with a remote Cloudflare DNS server, using both Ethernet and Wi-Fi connections.
The results, depicted in
\cref{f:nxdomain-histogram}, show that all the timers are fairly stable, with
little jitter.

For an Ethernet connection to a local DNS server, the timer
resolution is about 2\,ms, which \citet{ShustermanKHMMOY19} report is high
enough for the basic cache occupancy channel. A local server over Wi-Fi gives a
resolution of about~9\,ms, and the Cloudflare server provides a resolution of
roughly~70\,ms, for both Ethernet and Wi-Fi. While these resolutions are
unlikely to be suitable for the basic cache occupancy attack,
\citet{ShustermanKHMMOY19} show that sweep counting works well with the 100\,ms
timer of the Tor Browser.

\begin{figure*}[htb]
  \centering
  \begin{subfigure}[b]{0.30\textwidth}
    {\footnotesize\centering
      \begin{tikzpicture}
        \tikzstyle{class} = [rectangle, text centered, draw=black, fill=yellow!20]
        \node [class] (WPoT) at (0,2) {\shortstack{\strut Web Page\\\strut on Target}};
        \node [class] (IDS) at (4,2) {\shortstack{\strut Innocent\\\strut DNS Server}};
        \node [rectangle, draw=black, minimum width=0.3cm, minimum height=2cm] (l-WPoT) at (0,0) {};
        \node [rectangle, draw=black, minimum width=0.3cm, minimum height=2cm] (l-IDS) at (4,0) {};
        \draw [dotted] (WPoT) -- (l-WPoT);
        \draw [dotted] (IDS) -- (l-IDS);
        \draw [dotted] (l-WPoT) -- (0,-1.3);
        \draw [dotted] (l-IDS) -- (4,-1.3);
        \draw [-triangle 45] (0.15,0.8) -- (3.85,0.8) node [midway] {\shortstack{Resolve Non-Existent \\ Domain}};
        \draw (0.15,0.15) -- (1, 0.15) -- (1,-0.15) node [midway, right] {Probe Cache} [-triangle 45] -- (0.15,-0.15);
        \draw [-triangle 45, dashed] (3.85, -0.8) -- (0.15, -0.8) node [midway, above] {NXDOMAIN Err};
      \end{tikzpicture}
    }
    \caption{DNS Racing\label{f:dnsracing-interaction}}
  \end{subfigure}
  \hspace{0.04\textwidth}
  \begin{subfigure}[b]{0.3\textwidth}
    {\footnotesize\centering
      \begin{tikzpicture}
        \tikzstyle{class} = [rectangle, text centered, draw=black, fill=yellow!20]
        \node [class] (WPoT) at (0,2) {\shortstack{\strut Web Page\\\strut on Target}};
        \node [class] (MWS) at (3,2) {\shortstack{\strut Malicious\\\strut WebSocket Server}};
        \node [rectangle, draw=black, minimum width=0.3cm, minimum height=2cm] (l-WPoT) at (0,0) {};
        \node [rectangle, draw=black, minimum width=0.3cm, minimum height=2cm] (l-MWS) at (3,0) {};
        \draw [dotted] (WPoT) -- (l-WPoT);
        \draw [dotted] (MWS) -- (l-MWS);
        \draw [dotted] (l-WPoT) -- (0,-1.3);
        \draw [dotted] (l-MWS) -- (3,-1.3);
        \draw [-triangle 45] (0.15,0.8) -- (2.85,0.8) node [midway, above] {Send Short Packet};
        \draw (0.15,0.30) -- (1, 0.30) -- (1,-0.00) node [midway, right] {\shortstack{Search in\\String}} [-triangle 45] -- (0.15,-0.00);
        \draw [-triangle 45] (0.15,-0.8) -- (2.85,-0.8) node [midway, above] {Send Short Packet};
        \draw (3.15,0.80) -- (3.7, 0.80) -- (3.7,0.60) node [midway, right] {\shortstack{Log Start\\Time}} [-triangle 45] -- (3.15,0.60);
        \draw (3.15,-0.80) -- (3.7, -0.80) -- (3.7,-1) node [midway, right] {\shortstack{Log End\\Time}} [-triangle 45] -- (3.15,-1);
      \end{tikzpicture}
    }
    \caption{String and Sock\label{f:sns-interaction}}
  \end{subfigure}
  \hspace{0.04\textwidth}
  \begin{subfigure}[b]{0.3\textwidth}
    {\footnotesize\centering
      \begin{tikzpicture}
        \tikzstyle{class} = [rectangle, text centered, draw=black, fill=yellow!20]
        \node [class] (WPoT) at (0,2) {\shortstack{\strut Web Page\\\strut on Target}};
        \node [class] (MDS) at (3,2) {\shortstack{\strut Malicious\\\strut DNS Server}};
        \node [rectangle, draw=black, minimum width=0.3cm, minimum height=2cm] (l-WPoT) at (0,0) {};
        \node [rectangle, draw=black, minimum width=0.3cm, minimum height=2cm] (l-MDS) at (3,0) {};
        \draw [dotted] (WPoT) -- (l-WPoT);
        \draw [dotted] (MDS) -- (l-MDS);
        \draw [dotted] (l-WPoT) -- (0,-1.3);
        \draw [dotted] (l-MDS) -- (3,-1.3);
        \draw [-triangle 45] (0.15,0.8) -- (2.85,0.8) node [midway, above] {Resolve Domain};
        \draw (0.15,0.30) -- (1, 0.30) -- (1,-0.00) node [midway, right] {\shortstack{Search in\\String}} [-triangle 45] -- (0.15,-0.00);
        \draw [-triangle 45] (0.15,-0.8) -- (2.85,-0.8) node [midway, above] {Resolve Domain};
        \draw (3.15,0.80) -- (3.7, 0.80) -- (3.7,0.60) node [midway, right] {\shortstack{Log Start\\Time}} [-triangle 45] -- (3.15,0.60);
        \draw (3.15,-0.80) -- (3.7, -0.80) -- (3.7,-1) node [midway, right] {\shortstack{Log End\\Time}} [-triangle 45] -- (3.15,-1);
      \end{tikzpicture}
    }
    \caption{CSS \pp\label{f:css-pp-interaction}}
  \end{subfigure}
  \caption{Interaction diagrams for attacks.}
\end{figure*}
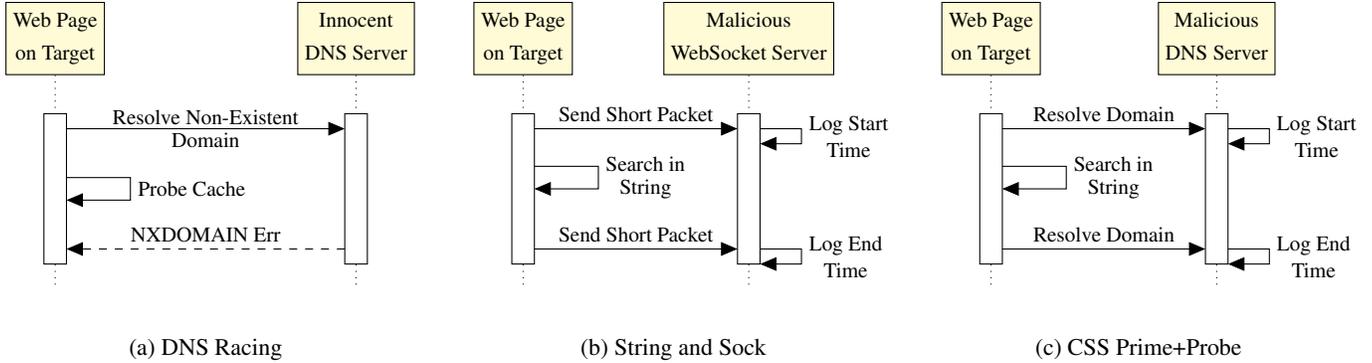

\parhead{Exploiting DNS for Cache Attacks.}
\cref{f:dnsracing-interaction} shows how to use the DNS response as a timer. 
As illustrated in the figure, the attacker first sets the \texttt{src}
attribute of an image to a non-existent domain, in
causing the operating system to access a remote DNS server for address resolution.
The attacker then starts the cache probe operation, creating a race
between the probe and the asynchronous report of the DNS error.
When the asynchronous error handling function is called after name resolution fails, 
the attacker can determine whether the cache probing operation was faster or slower than the network round-trip time.
 Alternatively,
when the DNS round-trip time is large, the attacker can repeat the probe step,
counting the number of probes before the DNS error is reported.
We note that the attack generates a large number of DNS requests.
Such anomalous traffic may be detected by intrusion detection systems and blocked by the firewall.

\subsection{String and Sock}\label{sec:stringandsock}
Another commonality feature of most  microarchitectural attacks in browsers,
including our DNS racing attack,  is the use of arrays~\cite{gruss2015practical, LippGSBMM17, GenkinPTY18}.
Consequently, the use of arrays is often assumed essential for performing
cache attacks in browsers and suggested countermeasures aim
for hardening arrays against side channels, while maintaining their functionality~\cite{SchwarzLG18}.
To refute this assumption, in this section we investigate a weaker attack
model, in which the attacker cannot use \js arrays and similar data structures.

\parhead{Exploiting Strings.}
Instead of using \js arrays, our \emph{String and Sock} attack uses
operations on long HTML strings. Specifically, we initialize a very long string
variable covering the entire cache. Then, to perform a cache contention
measurement, we use the standard \js \texttt{indexOf()} function to search for a
short substring in this long text.
We make sure that the substring we search for does not appear within the long
string, thus ensuring that the search scans all of the long string. Because the
length of the long string is the same as the size of the LLC, the scan
effectively probes the cache without using any \js array object. To measure the
duration of this probe operation, we take advantage of an external
WebSockets~\cite{RFC6455} server controlled by the attacker.

\parhead{Socket-Based Time Measurement.}
\cref{f:sns-interaction} shows how the String and Sock method operates. 
The attacker first sends a short packet to a cooperating WebSockets server. 
Next, the attacker performs a string search operation which is known to fail. 
As this search scans the entire string before failing, it has the side effect of probing the entire LLC cache. 
Finally, the attacker sends a second short packet to the cooperating WebSockets server. 
The server calculates the timing difference between the first and second packets, 
arriving at an estimate of the time taken to probe the cache.

\parhead{String and Sock in Chrome.}
We find that Chrome allocates three bytes for each character. As we would like our string to occupy the machines entire last level cache, we allocate different string lengths for each architecture considered in this paper. In particular, we use 2\,MiB strings for our Intel machines that feature a 6\,MiB LLCs, 3\,MiB strings for our AMD machines (4x16\,MiB LLCs), 1.5 \,MiB strings for our Samsung phones (8\,MiB LLC), and 2\,MiB strings for our Apple machines (12\,MiB LLCs on performance cores).
 We also note that Chrome caches results of recent searches. To
bypass this caching, for each search we generate a small fresh sequence of emojis and
search for it.
With the long string consisting only of ASCII characters, it is guaranteed not to contain any emojis.

\subsection{CSS \pp}\label{sec:csspp}
Our final attack, \emph{CSS \pp} targets an even more restricted setting,
in which the browser does not support \js or any other scripting language,
for example due to the NoScript extension~\cite{noscript}.
CSS \pp  only uses plain HTML and Cascading Style Sheets
(CSS) to perform a cache occupancy attack, without using \js at all.

\parhead{CSS \pp Overview.} At a high level, CSS \pp
builds on the String-and-Sock approach, and like it relies on string
search for cache contention and an attacker-controlled server for timing, see \cref{f:css-pp-interaction}.
Here, the attacker first includes in the CSS an element from an attacker-controlled domain, forcing DNS resolution. 
The malicious DNS server logs the time of the incoming DNS request.
The attacker then designs an HTML page that evokes a string search from CSS,
effectively probing the cache.
This string search is followed by a request for a CSS element that requires DNS resolution from the malicious server. 
Finally, the time difference between consecutive DNS requests corresponds to the time it takes
to perform the string search, which as described above is a proxy for cache contention.

\parhead{CSS \pp Implementation.}
\cref{fig:css_pp0} shows a code snippet implementing CSS \pp, using CSS \emph{Attribute Selectors} to perform the attack. 
Specifically, \cref{line:longclass} defines a
\texttt{div} with a very long class name (two million characters). This \texttt{div}
contains a large number of other \texttt{div}s, each with its own ID
(\crefrange{line:divstart}{line:divend}). The page also defines a style for each
of these internal \texttt{div}s (\crefrange{line:stylestart}{line:styleend}).
Each of these matches the IDs of the internal and external \texttt{div}, and
uses an attribute selector that searches for a substring in the external
\texttt{div}. If not found, the style rule sets the background
image of the element some URL at an attacker-controlled domain.

\begin{figure}[htb]
  \centering
  \begin{lstlisting}[style=web,escapechar=|]
<head>
	<style>
		#pp:not([class*='vukghj']) #s0 {background-image: url("https://kxdfvcgx.attack.com");}|\label{line:stylestart}|
  [...]
		#pp:not([class*='vatwjo']) #s9999 {background-image: url("https://bwpqxunq.attack.com");}|\label{line:styleend}|
	</style>
</head>
		<body>
			<div id="pp" class="AA...A">|\label{line:longclass}|
				<div id="s0">X</div>|\label{line:divstart}|
[...]
				<div id="s9999">X</div>|\label{line:divend}|
			</div>
		</body>
  \end{lstlisting}
\vspace{-1em}
  \caption{Simplified version of CSS-based \pp.}
  \label{fig:css_pp0}
\end{figure}

When rendering the page, the browser first tries to render the first internal
\texttt{div}. For that, it performs a long search in the class name, effectively
probing the cache occupancy. Having not found the substring, it sets the
background image of the \texttt{div}, resulting in sending a request to the
attacker's DNS server. The browser then proceeds to the next internal
\texttt{div}. As a result of rendering this page, the browser sends to the
attacker a sequence of DNS requests, whose timing depends on the cache
contention.

\begin{table*}[t] {
  \begin{center}
    \begin{tabular}{@{}lccccccccc@{}}
      \toprule
      & \multicolumn{4}{c}{Top-1 Accuracy (\%)} &&  \multicolumn{4}{c}{Top-5 Accuracy (\%)} \\
      \cmidrule{2-5}
      \cmidrule{7-10}
		       & Intel  & AMD Ryzen 9 & Apple & Samsung && Intel  & AMD Ryzen 9 & Apple & Samsung \\
                       
      Attack Technique & i5-3470 &  3900X & M1 & Exynos 2100 && i5-3470 &  3900X & M1 & Exynos 2100 \\
      \midrule
      Cache Occupancy & 87.5 & 69.1 & 89.7 &   84.5       && 97.0 & 91.4   & 97.8 &   95.3 \\
      Sweep Counting  & 45.8 & 54.9 & 90.5 &   69.7       && 74.3 &   82.9 & 98.1 &   91.5 \\
      DNS Racing      & 50.8 &  5.4 & 48.2 &   5.8       && 78.5 &   16.3 &   83.5 &   37.1 \\
      String and Sock & 72.0 & 53.9 & 90.6 &   60.2       && 90.6 & 85.5   & 97.9 &   85.5 \\
      CSS \pp        & 50.1 &   ---   & 15.7 &   ---       && 78.6 &  ---     &   32.6 &   --- \\
      \bottomrule
    \end{tabular}
  \end{center}
}
\vspace{-1em}
    \caption{\label{t:fp-res-arch}Closed-world accuracy (percent)
	across different microarchitectures.}
\end{table*}
\begin{table}[htb] {\footnotesize
		
		\begin{centering}
			\begin{tabular}{@{}lrrrr@{}}
				\toprule
				&\multicolumn{1}{c@{}}{Intel}  & \multicolumn{1}{c@{}}{AMD Ryzen 9} & \multicolumn{1}{c@{}}{Apple} & \multicolumn{1}{c}{Samsung} \\
				
				Attack Technique &\multicolumn{1}{c@{}}{i5-3470}  & \multicolumn{1}{c@{}}{3900X} & \multicolumn{1}{c@{}}{M1}  & \multicolumn{1}{@{}c@{}}{Exynos 2100}\\
				\midrule
				Cache Occupancy  &   2.9\,ms  &    6.0\,ms &   6.3\,ms &     4.0\,ms \\
				Sweep Counting   & 100.0\,ms  &  100.0\,ms & 100.0\,ms &   100.0\,ms \\
				DNS Racing       &  20.3\,ms  &    1.8\,ms &   7.2\,ms &     2.9\,ms \\
				String and Sock  &   1.5\,ms  &    2.9\,ms &   2.6\,ms &     2.5\,ms \\
				CSS \pp          &   0.3\,ms  &    6.7\,ms &   0.3\,ms &    33.8\,ms \\
				
				\bottomrule
			\end{tabular}
		\end{centering}
	}\vspace{-1em}
			\caption{\label{t:fp-tmpres}Temporal accuracy of attack techniques across different microarchitectures.}
\end{table}

\subsection{Empirical Results}\label{s:results}
We now present the classification results of the attacks described in this section
across different CPU architectures.
\cref{t:fp-res-arch} summarizes the accuracy of the most likely prediction
of the classifier (Top-1), as well as the likelihood that the correct answer is
one of the top 5 results (Top-5). Finally, \cref{t:fp-tmpres} shows the temporal resolution of each measurement method,  calculated as the time it takes to capture the entire trace, divided by the
number of points in the trace.

\parhead{Cache Occupancy.}
This method uses JavaScript code both to iterate over the eviction buffer, and to measure time.
The JavaScript code goes iterates over the buffer using the technique of Osvik et~al.\xspace~\cite{OsvikST06} to avoid triggering the prefetcher, and is written to prevent speculative reordering from triggering the timing measurement before the eviction is completed. 
As can be seen from the results, this approach provides good accuracy on all of the targets we evaluated, obtaining a top-5 accuracy of over 90\% across all platforms.

\parhead{Sweep Counting.} This method is designed for situations with lower clock resolution, but still uses JavaScript both for cache eviction and for timing measurement. 
As the results show, this added limitation translates to a loss in accuracy for most targets, with the Apple M1 target the least affected by the reduced timer resolution.

\parhead{DNS Racing.} This method uses JavaScript for cache eviction, but switches to the network for timing measurements. 
This added limitation translates to a loss in accuracy for most targets, largely due to the added jitter of the network.
The targets most severely affected by the added jitter were the ARM-based mobile phones, which were connected to the network using a wireless link, and the AMD devices, which were located in a third-party data center whose network conditions were beyond our direct control. We hypothesize that these networking circumstances led to jitter related to DNS responses, causing the severe loss of accuracy for these targets.

\parhead{String and Sock.} This is the first method which repurposes the browser's string-handling code for cache eviction. 
Unlike the adversary-controlled code used for mounting the cache occupancy attack described earlier, this third-party code naturally makes no attempt to trick the processor's cache management heuristics, and, as such, we expected it to have lower performance than the JavaScript-based code.

As we see, this was indeed the case for the Intel, AMD and Samsung targets. 
The Apple M1 target, on the other hand, did not encounter a loss in accuracy.
It seems that, on this target, na\"ively accessing a large block of memory is an  efficient way to evict the cache, and more advanced approaches for tricking the processor's prefetcher are not necessary.

\parhead{CSS \pp.}
As CSS \pp requires no \js, we test this attack in the presence of the NoScript \cite{noscript}
extension, applying the countermeasure only to our attacker website. As our
attack does not use \js at all, NoScript does nothing to prevent it. The accuracy
we obtained using this attack was comparable to the one obtained by the String
and Sock attack, showing that there is no need for \js, or any other mobile
code, to mount a successful side-channel attack. 

When running this attack on the Intel target, the accuracy is similar to DNS racing, which uses \js for cache evictions. 
On the M1 target, there was still a significant amount of data leaked by the attack, but the accuracy was less than the DNS racing attack.
On the ARM and AMD targets, we are unable at the present to extract any meaningful data using this method. As our CSS \pp also relies on DNS packets, we conjecture that this is due 
to the network conditions of the devices under test, or due to particular aspects of the micro-architecture of these devices which make cache eviction less reliable.

\parhead{Architectural Agnosticism.}
As the results show, we were able to mount our side-channel attack across a large variety of diverse computing architectures.
In particular, the Intel, AMD, ARM and Apple target architectures all incorporate different design decisions concerning different cache sizes, cache coherency protocols and cache replacement policies, as well as related CPU front-end features such as the prefetcher.
The reduced requirements of our attack made it immediately applicable to all of these targets, with little to no tuning of the attack's parameters, and without the need of per-device microarchitectural reverse engineering.

\parhead{Attacking Apple's M1 Architecture.}
To the best of our knowledge, this is the first side-channel attack on Apple's M1 CPU. 
The memory and cache subsystem of this new architecture have never been studied in detail, leading one to hope for a ``grace period'' where attackers will find this target difficult to conquer.
As this work shows, the novelty and obscurity of this new target do little to protect it from side-channel attacks.
The M1 processor is rumored to toggle between two completely different memory ordering mechanisms, based on the program it is executing.
Another noteworthy outcome from the M1 evaluation is that both the native arm64 binary of Chrome, as well as the standard MacOS Intel x64 Chrome binary running under emulation, were vulnerable to the attacks we described here.

Finally, observing \cref{t:fp-res-arch}, it can be seen that our attacks are, somewhat ironically, more effective on M1 architecture, than they are on other architectures, including the relatively well studied Intel architecture.  
Intel x86 CPUs are known to have advanced cache replacement and prefetcher policies, which are have been shown in other works to anticipate and mitigate the effect of large memory workloads on cache performance~\cite{DBLP:journals/micro/QureshiJPSE08, DBLP:conf/uss/BriongosMM020, DBLP:conf/dac/WangQAK19}.
We hypothesize that the M1 architecture makes use of less advanced cache heuristics, and that, as a result, the simplistic memory sweeps our attack performs are more capable of flushing the entire cache on these devices than they are on the Intel architecture.
This in turn results in a higher signal-to-noise ratio for the attack on these newer targets, and therefore in a higher overall accuracy.

\section{Attack Scenarios}\label{sec:scenarios}
We now turn our focus to a deeper investigation of the two new attacks we present,
String and Sock and CSS \pp, on the Intel targets.
\cref{t:variations-summary} provides a summary of the results discussed in this section.

\begin{table}[htb]
  \small\centering
  \begin{tabular*}{\columnwidth}{@{}l@{\extracolsep{\fill}}rr@{}}
    \toprule
    Attack Scenario    & String and Sock & CSS \pp  \\
    \midrule
    Closed World       & 74.5±1.6  & 48.8±1.6  \\
    Open World         & 80.2±1.1  & 60.9±1.4  \\
    Artificial Jitter  & 40.6±1.9  & 26.6±1.4    \\
    Tor Browser        & 19.5±8.7  & ---       \\
    DeterFox           & ---       & 65.7±1.2  \\
    %\midrule
    %Base Rate & 1 & 1   \\
    \bottomrule
  \end{tabular*}
  \caption{\label{t:variations-summary}Attack accuracy (\%) with 95\% confidence intervals.}
\end{table}

\subsection{Closed World Evaluation on Newer Intel Architectures}\label{s:closed-world}
We begin by reproducing the closed world methodology and
the results of \cref{s:cache-attacks-without-timers}
albeit on a newer Intel processor.
Specifically, we perform the experiments
on an Apple Macbook Pro featuring an
Intel Core i5-7267 CPU with a 4\,MiB last-level cache, and 16\,GiB memory,
running macOS 10.15 and Chrome version~81.
Despite the microarchitectural changes across 4 CPU generations and the different cache size,
the results are very similar to those achieved on the older i5-3470
(72.0±1.3\% for String and Sock and  50.1±2.3 for CSS \pp), with the difference being well inside the statistical confidence levels. We thus argue that our results transfer across a verity of Intel architectures.

\subsection{Open-World Evaluation}
A common criticism of closed-world evaluations is that the attacker is assumed to know the
complete set of websites the victim might visit, allowing the attacker to prepare and train classifiers for these websites~\cite{JuarezAADG14}.
For a more realistic scenario,  we follow the methodology proposed by \citet{PanchenkoNZE11} and perform an open-world evaluation, collecting 5000 traces of different websites used in \cite{RimmerPJGJ18},
in addition to the Alexa Top 100 websites collected in the closed-world setting.
We use the same data collection setting as for the closed-world collection. (See \cref{s:closed-world}.)

Here, the attacker's goal in this setting is to first detect if the victim visits one of the Alexa Top 100 sites, and secondly to identify the website if it is indeed in the list.
We note that in this case, a naive classifier can always claim that the site is not one of the Alexa Top 100, achieving a base rate of 30\%, resulting in slightly higher accuracy scores for any classifier.

In this open-world setting, the String and Sock and CSS \pp attacks obtain accuracy results of 80\% and 61\%, respectively.
The data in this setting is unbalanced --
there are more traces from ``other'' web sites than from each of the Alexa Top 100 sites.
For such data, the F1 score may be more representative than accuracy.
The F1 scores are
67\% and 45\%, for String and Sock and CSS \pp, respectively.
These are similar to those of the closed-world settings (70\% and 48\%).
We can therefore conclude that our attacks are as effective in the open-world as in the closed-world setting.

\begin{figure*}[htb]
  \centering
  \begin{subfigure}[b]{0.32\textwidth}
    \includegraphics[trim={0cm 0cm 0cm 0cm}, width=\linewidth]{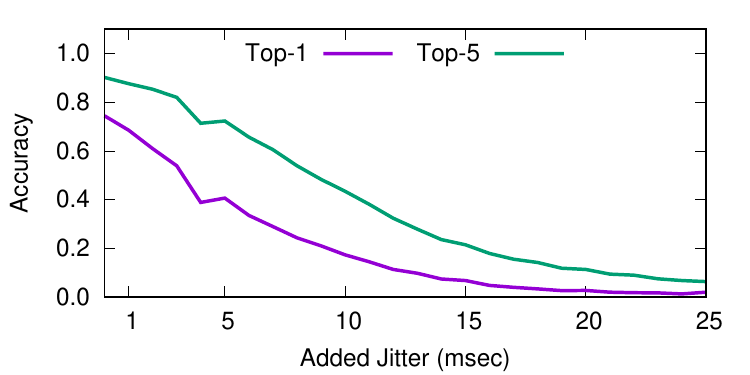}
    \caption{String and Sock
      \label{f:sns-jitter}}
  \end{subfigure}
  \begin{subfigure}[b]{0.32\textwidth}
    \includegraphics[trim={0cm 0cm 0cm 0cm}, width=\linewidth]{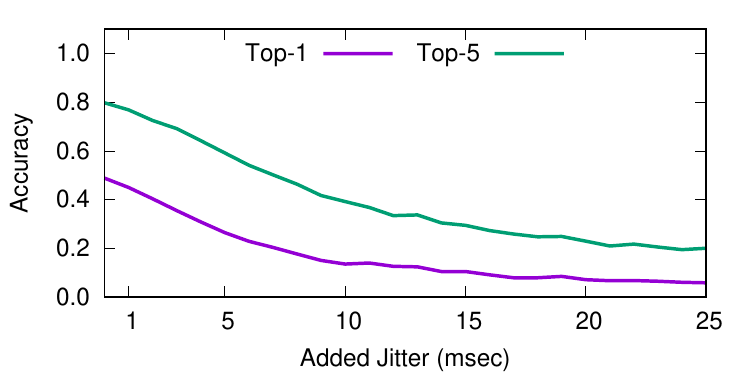}
    \caption{CSS \pp
      \label{f:css-jitter}}
  \end{subfigure}
  \begin{subfigure}[b]{0.32\textwidth}
    \includegraphics[trim={0cm 0cm 0cm 0cm}, width=\linewidth]{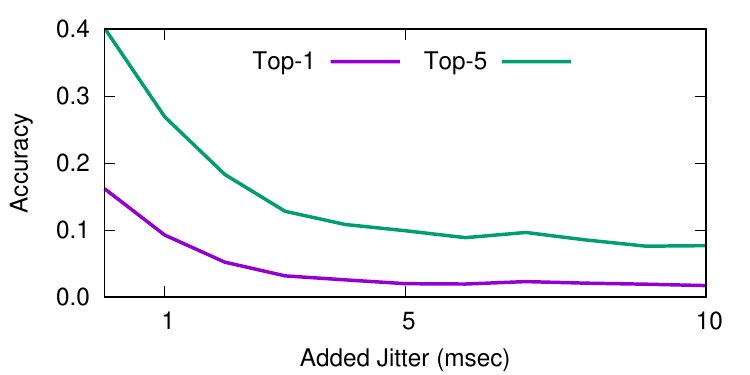}
    \caption{DNS Racing (note different scale)
      \label{f:dns-jitter}}
  \end{subfigure}
  \caption{Attack classifiers performance with additional jitter.}
  \label{f:jitter}
\end{figure*}

\subsection{Robustness to Jitter\label{s:robustness-to-jitter}}

As DNS racing, String and Sock, and CSS \pp use an external server for time measurement, these techniques are inherently sensitive to jitter naturally present on the network between the victim and the web server.

\begin{figure}[htb]
	\includegraphics[trim={0cm 0cm 0cm 0cm}, width=\linewidth]{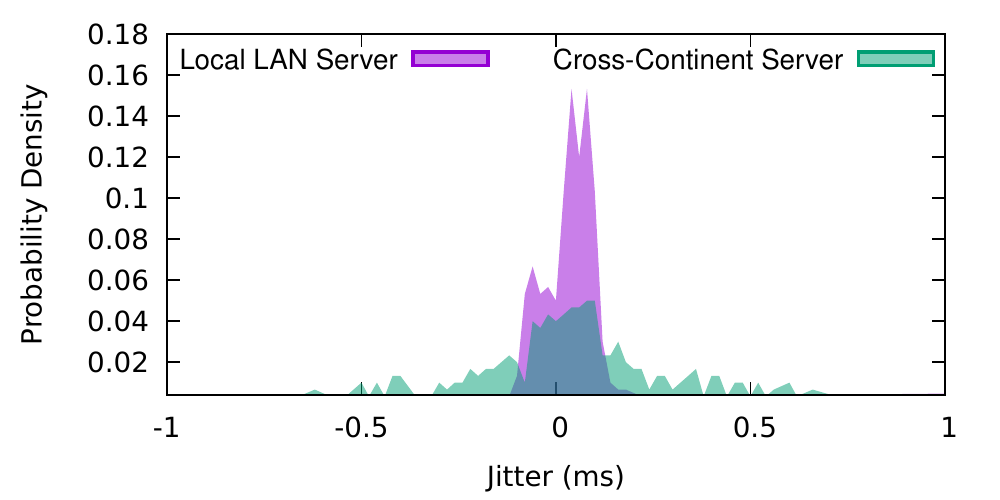}
	\caption{Measured Jitter of the WebSockets server response.
		\label{f:string-n-sleep-histogram}}
\end{figure}

\parhead{Measuring Network Jitter.} We measure the network jitter in two scenarios. First, we perform a local measurement, where the target and an attacker-controlled WebSockets server are located on the same institutional network at Ben Gurion University, Israel.  Next, we also perform an inter-continental measurement, where the attacker is located in Israel, while the server is located in the United States (University of Michigan). \cref{f:string-n-sleep-histogram} shows the distribution of the jitter observed while sending 100 packets per second 
for 30 seconds to the WebSockets servers. We find that the jitter in the local network has a 
standard deviation of 0.17\,ms, whereas the jitter to the cross-continent server has standard deviation of 0.78\,ms.

\parhead{Evaluating Robustness to Jitter.}
Having established the typical jitter between the target and the external server, we now evaluate the robustness of our techniques to various levels of jitter. To that aim, 
 we artificially inject different amounts of jitter to the closed-world dataset of \cref{s:closed-world}.
The jitter is injected by adding random noise to the timing of the monitored events.
This noise is selected at random from a normal distribution with a mean zero
and a standard deviation that varies from 1 to 25 milliseconds,
with higher standard deviation corresponding to larger jitter.

As \cref{f:jitter}
shows, both the String and Sock and the CSS \pp attacks
still retain most of their accuracy even if the jitter is an order of magnitude
larger than the ones we measured on a real network.
We finally note that the DNS Racing attack is more sensitive to added jitter,
as it relies on a binary race condition to determine timing.

\section{Analysis of an API-based Defense}
\label{sec:attacking-crz}
Having established the efficacy of our techniques on various microarchitectures, in this section we evaluate our attacks in the presence of increasing levels of browser hardening. 

To that aim,  we make use of \crz~\cite{SchwarzLG18}, a Chrome extension that supports per-website restrictions on \js browser API features.
We begin by  presenting an overview of \crz's \js implementation and security objectives, focusing on a subset of \crz's features which are relevant to this work.
We next describe how we modified \crz to offer more comprehensive protection, at the cost of usability and performance. Finally, we show that even with these modifications, \crz is unable to offer side channel protections against the techniques presented in this paper. Unless stated otherwise, we use the current version at \crz's Git
repository.\footnote{\url{https://github.com/IAIK/ChromeZero} commit
  \texttt{fee8adc6c8fce9dd1ab62d7ff8f0697b44a188c1}}

\begin{table*}[htb]
	\begin{centering}
		\resizebox{\linewidth}{!}{
			\begin{tabular}{@{}lllll@{}}
				\toprule
				Policy Level        & \pollow     & \polmed                  & \polhigh                & \polpara                  \\
				\midrule
				Memory Addresses    & Buffer ASLR & Array preloading         & Non-deterministic array & Array index randomization \\
				Timer manipulation  & Ask User    & Low-resolution timestamp & Fuzzy time              & Disabled                  \\
				Multithreading      & ---         & Message delay            & WebWorker polyfill      & Disabled                  \\
				Shared Array Buffer & ---         & Slow SharedArrayBuffer   & Disabled                & Disabled                  \\
				Sensor API          & ---         & Ask User                 & Fixed Value             & Disabled                  \\
				\bottomrule
			\end{tabular}
		}
		\caption{\label{tab:crz-policies}Defense techniques used in each  \crz Policy Level.}
		\par\end{centering}
\end{table*}

\subsection{\crz Overview}
\crz implements a list-based access control policy, which dictates actions
to be taken when a website invokes a \js function or accesses an object
property. When an access is detected, \crz either allows the access, modifies it,
or completely blocks the access
based on the policy chosen for the particular website.\footnote{\crz
  currently only supports a global protection policy that can be changed
  but applies to all websites.}
\crz also supports the option of asking the user about the action to take.

\parhead{Default Policies.}
\crz offers five preset protection policies for the
user to choose from: \poloff, \pollow, \polmed, \polhigh, and \polpara.
\footnote{The \crz extension uses the name ``Tin Foil Hat'' for \polpara. We
  stick to the naming in \citet{SchwarzLG18}.} As it progresses through protection
policy levels, \crz  makes increasingly severe restrictions on \js capabilities
and resources, including blocking them altogether. \cref{tab:crz-policies}
summarizes which capabilities and resources are available at each protection
level.

\parhead{Performance.}
\citet{SchwarzLG18} claim that \crz blocks all of the building blocks required
for successful side-channel attacks, including high resolution timers, arrays
and access to hardware sensors. Moreover, they claim that \crz prevents many
known CVEs and 50 percent of zero-day exploits published since chrome 49.
Finally, \citet{SchwarzLG18} benchmark \crz's performance and perform a
usability study. They claim that \crz has an average overhead of 1.82\% at the
second-highest protection level (\polhigh) and that its presence is
indistinguishable to users in 24 of Alexa's Top 25 websites.

\begin{figure}
  \centering
  \includegraphics[width=6cm]{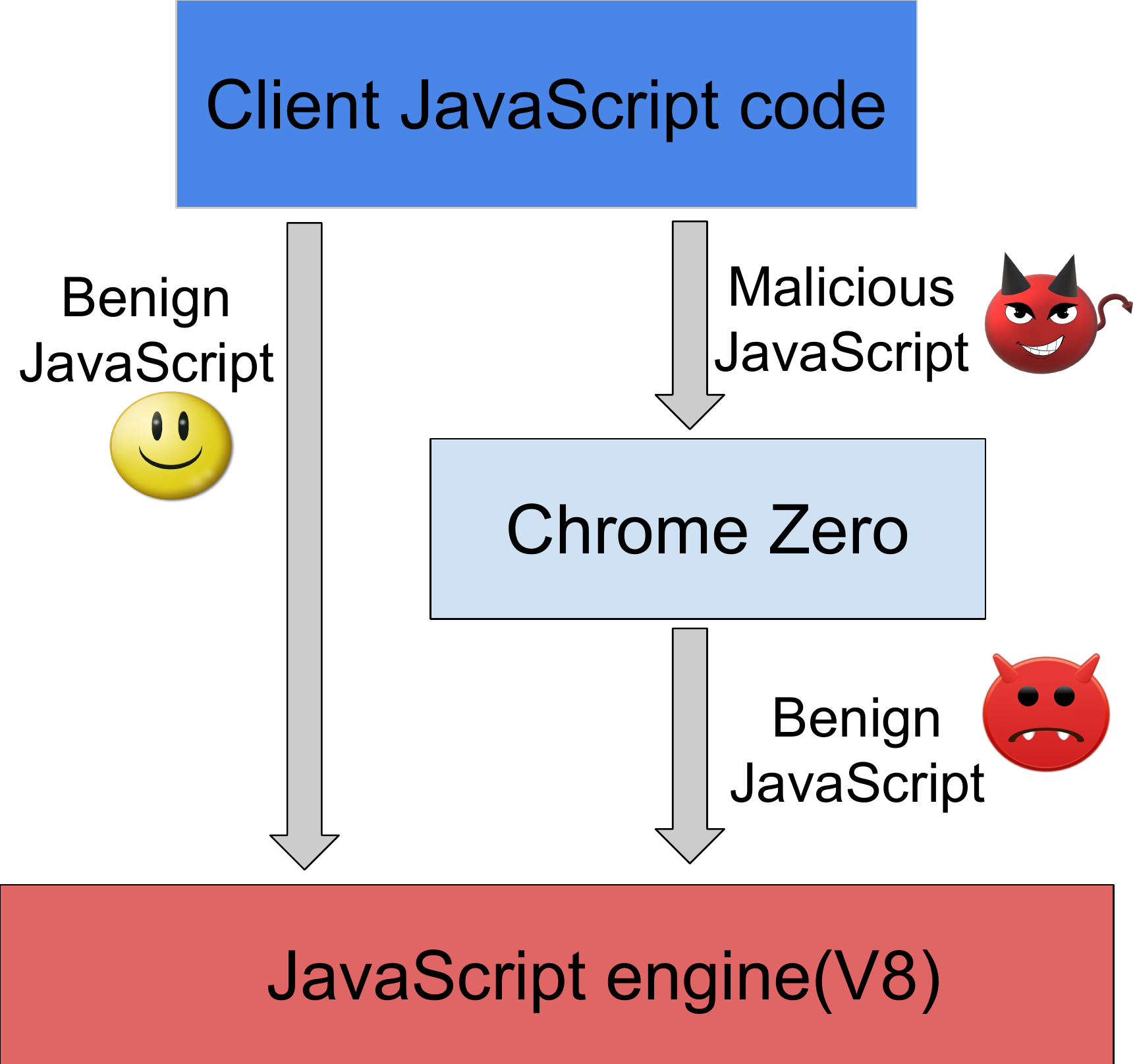}
  \caption{High-level concept of \crz}
  \label{fig:interception}
\end{figure}

\parhead{\crz's Access Control Implementation.}
To enforce security policies, \crz intercepts
\js API calls using Virtual Machine Layering. Specifically, \crz is implemented as \js code that is injected into a
web page when upon initialization. This injected code
wraps sensitive API functions, having the wrappers
implement actions specified by \crz's policy. \crz uses closures to ensure that the wrapper contains the only reference to the original API functions, thus ensuring that
websites do not trivially bypass its protection
(\cref{fig:interception}).

\parhead{Protecting Timers.}
Traditionally, microarchitectural side-channel attacks rely on having access to
a high-resolution timer, e.g.\ to distinguish cache hits from cache misses. This
includes attacks implemented in native code~\cite{OsvikST06, Percival05,
  YaromF14, GrussMW15, GrussSM15, LiuYGHL15, YaromGH16, GrasRBG18, AciicmezKS07}
as well as attacks in \js running inside the
browser~\cite{OrenKSK15,GrasRBBG17,SchwarzMGM17,GenkinPTY18}. As a
countermeasure for such attacks, Chrome's current implementation of
\texttt{performance.now()} already reduces timer resolution from nanoseconds to
microseconds and introduces a small amount of jitter. Although these mitigations protect against some high-resolution
attacks~\cite{OrenKSK15,GrasRBBG17,SchwarzMGM17}, microsecond-accurate timers still
provide sufficient resolution for other side-channel attacks from within \js~\cite{stone2013pixel, GrussMM15, SchwarzMGM17, gruss2015practical, van2015clock}.

To block attacks that exploit microsecond-accurate timers, \crz employs two main
strategies. At its \polmed protection policy, \crz applies a ``rounded
floor'' function, matching the 100\,ms resolution of the Tor Browser. While this
already prevents many attacks~\cite{SchwarzMGM17}, higher resolution
timers may still be constructed~\cite{kohlbrenner2016trusted,SchwarzMGM17,van2015clock}. Thus, at
higher protection levels, instead of using a simple ``rounded floor'' 100\,ms
timers, \crz follows the approach of \citet{vattikonda2011eliminating} and
fuzzes the timer measurements by adding random microsecond-level noise. Finally,
at its highest protection level, \crz disables timers altogether.

\parhead{Arrays.}
\citet{SchwarzLG18} identify that many side-channel attacks in
browsers~\cite{OrenKSK15,GenkinPTY18,gruss2015practical,GrussMM15,SchwarzMGM17,GrasRBBG17}
require some information about memory addresses. Typically, recovering the page
offset (least significant 12 of 21 bits of the address) facilitates the attacks.
Using this information the attacker
then analyzes the victim's behavior, deducing information about its control flow
and internal data.
\crz therefore applies several mitigations to \js array APIs.

More specifically, \crz's second-highest protection level introduces array
non-determinism, adding an access to a random element for each array access. The idea is that the random accesses  themselves force page faults, impeding the use of page faults as signals
for page boundaries. \citet{SchwarzLG18} argue that this method prevents eviction set construction~\cite{GrussMM15,OrenKSK15,SchwarzMGM17,GenkinPTY18,YaromGLLH15}, as
it interferes with the specific sequences required to construct an eviction set,
while adding noise to the timing information.

Next, \crz further deploys the buffer ASLR
policy, which shifts the entire buffer by a random offset.
This is achieved by intercepting the array constructors and access methods. To prevent page alignment, \crz increases the requested array size by 4\,KiB, and associates a random page offset with the  array. On array access, \crz adds the random offset to the requested array
index, thereby shifting the access by the random offset.

Finally, to protect the offset from being discovered, \crz attempts to
use the additional accesses to random elements to pre-load all the
array's memory pages into the cache, thus preventing attackers from detecting page boundaries
by looking for array elements which have an increased access time due to page
faults.

\parhead{Protecting Against Browser Exploits.} While not being a primary goal of  \crz, \citet{SchwarzLG18} argue that \crz is also
capable of protecting users against some browser exploits. To validate their
claim, they reproduced 12 CVEs 
listed in \cref{t:cvemitigation},
in the then-current Chrome \js engine, and found that \crz
prevents exploiting half of the CVEs. \citet{SchwarzLG18} attribute this
protection to the modification of \js objects in \crz, which breaks the CVE
exploit code.

\subsection{API Coverage}
\label{sec:crzproblems}
As stated above, \crz is essentially an interception layer, which intercepts the
critical \js API calls and subsequently directs them to the appropriate logic
based on the current website and protection policy. Thus, to guarantee security, it is
critical to ensure that malicious \js code cannot access the original API or otherwise bypass the \crz protections.

Our investigation of \crz demonstrated that API coverage in \crz leaves a lot
to be desired.
Specifically, we have identified multiple instances of APIs that are not protected by
\crz.  These include:
\begin{itemize}[nosep, leftmargin=*]
  \item \parhead{Delayed Extension Initialization.}
        The \crz extension initializes after the browser finishes constructing the
        Document Object Model (DOM) for the page.
        Consequently, \crz does not protect \js objects
        created before the DOM is constructed.
  \item \parhead{Missed Contexts.} \crz only applies its security policies in the
        context of the topmost page in each browser tab.
        It does not, however, protect code in sub-contexts of the page,
        including worker threads and iframes.
  \item \parhead{Unprotected Prototype Chains.}
        As we discuss in \cref{s:bgjstype}, properties of global objects
        may be inherited from their prototypes.
        Yet, while \crz does protect global objects, it fails to
        protect their prototype chains, allowing attackers to access the original \js API.
\end{itemize}

\parhead{Exploitation.}
We have exploited each of those omissions and demonstrated complete bypass
of \crz protections.
In most cases, such bypasses are fairly trivial.
As an example we show how we exploit unprotected prototype chains.

\begin{figure}[htb]
  \centering
  \includegraphics[width=8cm]{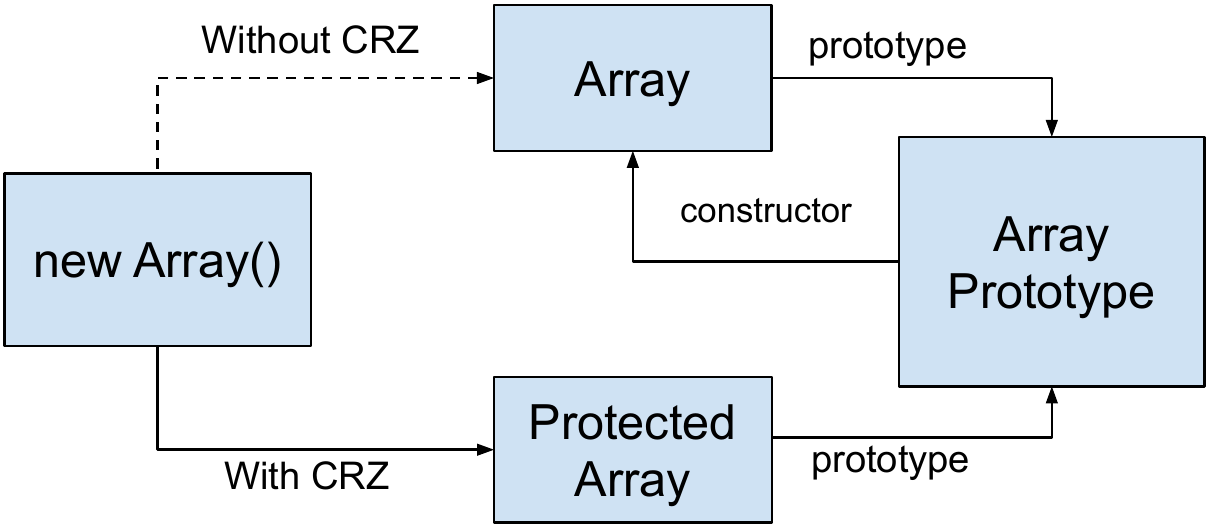}
  \caption{Object hierarchy with \crz.}
  \label{fig:prototype}
\end{figure}

\begin{figure}[htb]
  \centering
  \begin{lstlisting}[style=web,escapechar=|]
	let secureArray = new Array(10);|\label{line:secarray}|
	let secureTimer = performance.now();|\label{line:sectimer}|
	
	let insecureArray = new secureArray.__proto__.constructor(10);|\label{line:freearray}|
	let insecureTimer = performance.__proto__.now.call(performance);|\label{line:freetimer}|
	\end{lstlisting}
  \caption{Bypassing \crz defenses using prototypes.}
  \label{fig:prototype_attacks}
\end{figure}

\cref{fig:prototype} shows the object
hierarchy for \texttt{Array} with \crz (solid line) and without it (dotted line). The original unprotected
\texttt{Array} class can be accessed using the \texttt{Array} constructor method
of the prototype object. \cref{fig:prototype_attacks} shows a bypass of \crz
object protections, allowing the attacker to create original non-proxied \js
objects. \cref{line:secarray,line:sectimer} show the standard ways of creating
an array or getting the timer, both protected by \crz. In contrast,
\cref{line:freearray,line:freetimer} show how to use prototypes to achieve the
same functionality, bypassing \crz.

\begin{table}[htb]{\footnotesize
    \begin{centering}
      \begin{tabular}{@{}lllll@{}}
        \toprule
        CVE Number    & Chrome       & Original & Modified & Summary \\
                      & Version      & Exploit  & Exploit  &         \\
        \midrule
        CVE-2016-1646 & 49.0.2623.75 & \cmark   & \xmark   & \xmark  \\
        CVE-2016-1653 & 49.0.2623.75 & \cmark   & \cmark   & \cmark  \\
        CVE-2016-1665 & 50.0.2661.75 & \xmark   &          & \xmark  \\
        CVE-2016-1669 & 50.0.2661.75 & \cmark   & \xmark   & \xmark  \\
        CVE-2016-1677 & 50.0.2661.75 & \xmark   &          & \xmark  \\
        CVE-2016-5129 & 51.0.2704.84 & \xmark   &          & \xmark  \\
        CVE-2016-5172 & 51.0.2704.84 & \xmark   &          & \xmark  \\
        CVE-2016-5198 & 54.0.2840.71 & \xmark   &          & \xmark  \\
        CVE-2016-5200 & 54.0.2840.71 & \cmark   & \cmark   & \cmark  \\
        CVE-2016-9651 & 54.0.2840.71 & \cmark   & \xmark   & \xmark  \\
        CVE-2017-5030 & 54.0.2840.71 & \cmark   & \xmark   & \xmark  \\
        CVE-2017-5053 & 55.0.2883.75 & \xmark   &          & \xmark  \\
        \bottomrule
      \end{tabular} 
        \caption{CVE PoC exploits mitigated by \crz.  \cmark denotes a mitigated exploit while 
        \xmark denote unmitigated exploits. }
  \label{t:cvemitigation}
    \end{centering}
  }
\end{table}

\parhead{Evaluating \crz's CVE Protection.}
We also evaluate \crz's claimed protection against browser exploits.
We first reproduce the results of \citet{SchwarzLG18} finding that
\crz indeed prevents six of the 12 exploits they experiment with, 
see \cref{t:cvemitigation}. 
We note, however, that 
\crz only protects incidental properties of the
exploits rather than addressing the underlying vulnerabilities. In particular, 
we were able to easily modify the code of the blocked exploits
to avoid using features that \crz protects. Overall, we find that \crz only properly blocks two of the 12 CVEs evaluated by \citet{SchwarzLG18}. 

Next, we extend the evaluation of \crz to CVEs reported after the publication of  \citet{SchwarzLG18}. Here, we find that \crz blocks only four of the 17 exploits we were able to reproduce in Chrome, see \cref{t:newcvemitigation}. 
Yet again, some of these new exploits can be modified to not use the APIs blocked by \crz, 
allowing the exploits to work unhindered. In summary, we find that \crz was successfully able to  block only four CVEs out of the out of 29 reproducible Chrome CVEs.
%Finally,  we find a common root cause for \crz's ability to systematically mitigate these four exploits. 
%More specifically, we find that these use 
%
%for the four exploits we cannot modify to bypass \crz, we find that
%the cause is that the use of protected typed arrays prevents
%Chrome from compiling Web Assembly~\cite[``read the imports'']{wasm_standard}.
%Since the Web Assembly compiler is not invoked, the browser remains protected.

\begin{table}[htb]{\footnotesize
    \begin{centering}
      \begin{tabular}{@{}lllll@{}}
        \toprule
        CVE Number     & Chrome        & Original & Modified & Summary \\
                       & Version       & Exploit  & Exploit  &         \\
        \midrule
        CVE-2017-5070  & 58.0.3029.96  & \xmark   &          & \xmark  \\
        CVE-2017-5071  & 57.0.2987.133 & \xmark   &          & \xmark  \\
        CVE-2017-5088  & 57.0.2987.133 & \xmark   &          & \xmark  \\
        CVE-2017-5098  & ---           & ---      &          & ---     \\
        CVE-2017-5115  & 59.0.3071.86  & \xmark   &          & \xmark  \\
        CVE-2017-5116  & 60.0.3112.90  & \cmark   & \cmark   & \cmark  \\
        CVE-2017-5121  & 61.0.3163.79  & \xmark   &          & \xmark  \\
        CVE-2017-5122  & ---           & ---      &          & ---     \\
        CVE-2017-15399 & 62.0.3202.75  & \cmark   & \cmark   & \cmark  \\
        CVE-2017-15401 & ---           & ---      &          & ---     \\
        CVE-2018-6056  & 62.0.3202.75  & \cmark   & \xmark   & \xmark  \\
        CVE-2018-6061  & ---           & ---      &          & ---     \\
        CVE-2018-6064  & 62.0.3202.75  & \xmark   &          & \xmark  \\
        CVE-2018-6065  & 63.0.3239.108 & \xmark   &          & \xmark  \\
        CVE-2018-6092  & ---           & ---      &          & ---     \\
        CVE-2018-6106  & 63.0.3239.108 & \xmark   &          & \xmark  \\
        CVE-2018-6122  & ---           & ---      &          & ---     \\
        CVE-2018-6136  & 63.0.3239.108 & \xmark   &          & \xmark  \\
        CVE-2018-6142  & ---           & ---      &          & ---     \\
        CVE-2018-6143  & 63.0.3239.108 & \xmark   &          & \xmark  \\
        CVE-2018-6149  & 70.0.3538.77  & \cmark   & \xmark   & \xmark  \\
        CVE-2018-16065 & ---           & ---      &          & ---     \\
        CVE-2018-17463 & 68.0.3440.84  & \xmark   &          & \xmark  \\
        CVE-2019-5755  & 70.0.3538.77  & \xmark   &          & \xmark  \\
        CVE-2019-5782  & 70.0.3538.77  & \xmark   &          & \xmark  \\
        CVE-2019-5784  & ---           & ---      &          & ---     \\
        \bottomrule
      \end{tabular} 
        \caption{\crz mitigation of post-publication CVEs.
          \cmark and \xmark denote mitigated and unmitigated exploits (respectively)  while --- denotes CVEs that we were unable to reproduce with available PoCs}
  \label{t:newcvemitigation}
    \end{centering}
  }
\end{table}

\subsection{Fixing and Re-evaluating \crz}\label{sec:crzevaluation}
\crz's failure to protect all of the \js API has implications beyond security.
Unprotected objects do not affect the usability or the performance of the browser.
To evaluate the impact of the approach on usability and performance,
we fix \crz to improve its API coverage.
Specifically, we set
\crz to initialize before any other script executes and to also apply to frames.
We further modify \crz to apply its interception to protected objects and all
the objects in their prototype chain. We do not protect Web Workers, hence our
analysis below may still understate the impact on usability and performance.
We further remove bypasses of array protections that apply to some hard-coded websites.
Specifically, \crz does not apply some array protections to YouTube and to Google Maps.%
\footnote{We note that without the bypass, YouTube does not play videos.
  We could not find any indication of
  this bypass in \citet{SchwarzLG18}, which we find odd given the use of
  YouTube in the usability evaluation.
  The \crz source code claims that the bypass is due to a bug in Chrome,
  however our root cause analysis shows that YouTube fails to play videos
  due to the type mismatch we discuss in this section.}

Finally,
\citet{SchwarzLG18} argue that \crz offers no noticeable impact on user experience
while only having a negligible performance cost. We test this claim with and
without our security fixes.

\parhead{Experimental Setup.}
We use a ThinkPad P50 featuring an Intel Core i7-6820HQ CPU, with 16 GiB of
memory, running Ubuntu version 18.04, with a Chrome 80 browser without any
extensions. We evaluate usability on Alexa's Top~25 USA websites, checking for
discernible differences in behavior.

\begin{table}[htb]
  \begin{centering}
    \footnotesize
    \begin{tabular}{@{}lllllll@{}}
      \toprule
       & \multicolumn{2}{c}{Level} && & \multicolumn{2}{c}{Level} \\
      Domain      & \pollow & \polhigh && Domain & \pollow & \polhigh \\
      \cmidrule{1-3}
      \cmidrule{5-7}
      %\midrule
      google.com      & \xmark                 & \xmark                  &&
      youtube.com     & \xmark                 & \xmark                  \\
      amazon.com      & \xmark                 & \xmark                  &&
      facebook.com    & \xmark                 & \xmark                  \\
      yahoo.com       & \xmark                 & \xmark                  &&
      reddit.com      & \cmark                 & \xmark                  \\
      wikipedia.org   & \cmark                 & \cmark                  &&
      ebay.com        & \cmark                 & \cmark                  \\
      netflix.com     & \xmark                 & \xmark                  &&
      bing.com        & \cmark                 & \xmark                  \\
      office.com      & \cmark                 & \xmark                  &&
      live.com        & \cmark                 & \xmark                  \\
      myshopify.com   & \xmark                 & \xmark                  &&
      instructure.com & \xmark                 & \xmark                  \\
      twitch.tv       & \cmark                 & \xmark                  &&
      cnn.com         & \xmark                 & \xmark                  \\
      linkedin.com    & \cmark                 & \xmark                  &&
      instagram.com   & \cmark                 & \xmark                  \\
      espn.com        & \xmark                 & \xmark                  &&
      dropbox.com     & \xmark                 & \xmark                  \\
      intuit.com      & \cmark                 & \xmark                  &&
      nytimes.com     & \xmark                 & \xmark                  \\
      chase.com       & \cmark                 & \xmark                  &&
      tmall.com       & \cmark                 & \xmark                  \\
      \bottomrule
    \end{tabular} 
    \caption{Websites usability with \crz. \cmark  denotes working websites while 
    	\xmark denotes non-working websites.}
\label{t:usability}
  \end{centering}
\end{table}

\parhead{Usability Results.}
We first replicate the results of \citet{SchwarzLG18}, finding that an
unmodified \crz has no discernible impact on the usability of websites.
However, after fixing the issues identified in \cref{sec:attacking-crz},
we observe a significant impact on the usability of websites.
Even when setting \crz to the \pollow policy, less than half of the websites 
function without noticeable problems. 
At a higher
protection level, \polhigh,  only two websites function properly.
\cref{t:usability} summarizes the usability results for \crz on the 24
websites at the top of Alexa Top Websites (USA).
For a site to be considered ``perfectly working'', it needs to look identical to
the unprotected mode, display no additional error messages to the user, and have
working interaction features (scrolling, zooming, menus, search input, etc.)

\parhead{Strict Type Checking.}
Investigating the difference in website usability between the original and
modified \crz,  we find that forcing \crz to apply its policies before document
loading results in type mismatch exceptions while loading many \js-enabled web
sites.

The cause of the issue is that as part of applying its policies, \crz replaces
any \js object it protects with a proxy that masquerades as the original object.
Typically this does not cause any problems due to \js's use of ``duck typing'',
since replacing objects with the corresponding proxy objects is transparent to
most \js code, as long as the original object's properties are all supported.
However, the W3C standard~\cite{uint8array_set} dictates strict type checking
for many internal \js functions, especially for typed array objects. In this
case, passing a proxy object instead of the original object results in a type
mismatch exception from the browser's \js engine, causing the website's loading
to fail.

Unfortunately, fixing this issue turns out to be a non-trivial problem, as a significant portion of 
the \js environment is forced to strictly type check
its inputs. This goes well beyond the member functions of \texttt{TypedArray}s
and includes diverse \js libraries, such as, for example, the Web Crypto and Web Socket APIs.

\parhead{Estimating Performance Impact.}
While we do not claim to know an efficient method of automatically solving this problem for 
the entire \js API, we can efficiently solve the issue for specific functions through manual intervention, 
allowing us to benchmark the result. While we acknowledge that this does not produce a secure or even 
correct implementation, we argue that it nonetheless allows us to
measure a lower-bound of the performance impact that any \js zero
implementation must have.
To that aim, we enumerate all of the functions used by the JetStream~1.1 benchmark, and
manually implement fixes for functions that perform strict type checking. We note that only the
\texttt{set} and \texttt{subarray} methods for typed arrays need to be fixed, while
all other parts of the \js environment can remain unaltered.

\parhead{Benchmarking Performance}
For performance benchmarks we first try to reproduce the results of \citet{SchwarzLG18}.
We use the JetStream~1.1 benchmark to facilitate comparison with \citet{SchwarzLG18}.
We find a slight performance impact of 1.54\% when using an unmodified \crz. However,
when ensuring that \crz applies its protections correctly and applying the minimum
level of fixes for strict type checking we observe a performance impact of 26\%
in the latency benchmarks and 98\% in the throughput 
 benchmarks, as described in
\cref{app:detailed}.

\subsection{Bypassing Non-Deterministic Arrays}
\label{sec:non-det-arrays}

With the exception of speculative execution
attacks~\cite{KocherHFGGHHLM019,Lipp0G0HFHMKGYH18,BulckMWGKPSWYS18,CanellaB0LBOPEG19},
most microarchitectural side-channel attacks retrieve information about memory
access patterns performed by the victim.
For a language such as \js with no notion of pointers or
addresses, most attacks exploit the contiguous nature and predictable memory
layout of  arrays to reveal information about the least significant 12 or 21
bits of the addresses accesses by the victim~\cite{OrenKSK15, GrussMM15,
  SchwarzMGM17, GrasRBBG17}.

To prevent this leakage, \crz's second-highest protection level introduces array
non-determinism, performing a spurious access to a random array index whenever
the script accesses an array element. \crz further deploys the buffer ASLR
policy, which shifts the entire buffer by a random offset, thereby preventing
the attacker from obtaining page-aligned buffers. The main idea is to use the
random offset to deny the attacker from finding the array elements located on
page boundaries. To protect the offset from being discovered, \crz attempts to
use the additional accesses to random elements in order to pre-load all the
array's memory pages into the cache, thus preventing the attacker from
discovering the array elements which have an increased accesses time due to page
faults.

We now show how we can reliably recover the array elements corresponding to
page boundaries, despite \crz's use of buffer ASLR,
non-deterministic arrays, and fuzzy timers.

\parhead{Array Implementation in Chrome.}
Unlike their C counterparts, \js arrays are quite flexible and can be
extended~\cite{mdn_push}, shrunk~\cite{mdn_pop} and even have their
type changed~\cite{v8_kinds} at run-time. While the W3C standards require
browsers to support the extension and shrink APIs, the implementation of these
capabilities is left entirely to the browser vendors.

In Chrome's V8 \js engine, whenever an array is initialized, V8 allocates the
memory required for the array, along with an additional memory to support
insertion of more elements in $O(1)$ amortized time. However, after the
addition of enough elements, memory reallocation is eventually needed. Hence V8
allocates a new chunk of memory which is about $1.5 \times$ larger than the old
one, and frees the old one after copying the array's content to the new location.
The formula used by V8 to determine the size of the new memory buffer is

\begin{equation}
  \label{eqn:v8-resize}
  \mnewsize = \msize + \msize \gg 1 + 16,
\end{equation}
where $\gg$ is a bit-wise shift-right operation.

\begin{figure}[htb]
  \centering
  \begin{lstlisting}[style=web,escapechar=|]
let array = new Array();|\label{line:newarray}|
let times = new Array();
  
for(let i=0; i<10000000; i++){|\label{line:loop}|
  let start = performance.now();|\label{line:firsttimer}|
  array.push(0); |\label{line:push}|
  let delta = performance.now() - start;|\label{line:secondtimer}|
  times.push(delta);
}
\end{lstlisting}
  \caption{Measuring Array.push timings}
  \label{fig:array_extension}
\end{figure}

\parhead{Attack Methodology.}
We begin by measuring the timings of \texttt{Array.push} using the code
presented in \cref{fig:array_extension}.  We start with an empty array
\texttt{array} (\cref{line:newarray}). We then append data to the end of the
array using the \js \texttt{Array.push} method (\cref{line:push}). On every such
element addition we measure the time taken to add an element
(\cref{line:firsttimer,line:secondtimer}). While most of these additions are
fast, at the point where the memory allocated for the current size of
\texttt{array} is exhausted, V8 performs additional work by allocating new
memory using \cref{eqn:v8-resize} and copying the old content to the
newly-allocated space.

\begin{figure}[htb]
  \centering
  \includegraphics[width=\linewidth]{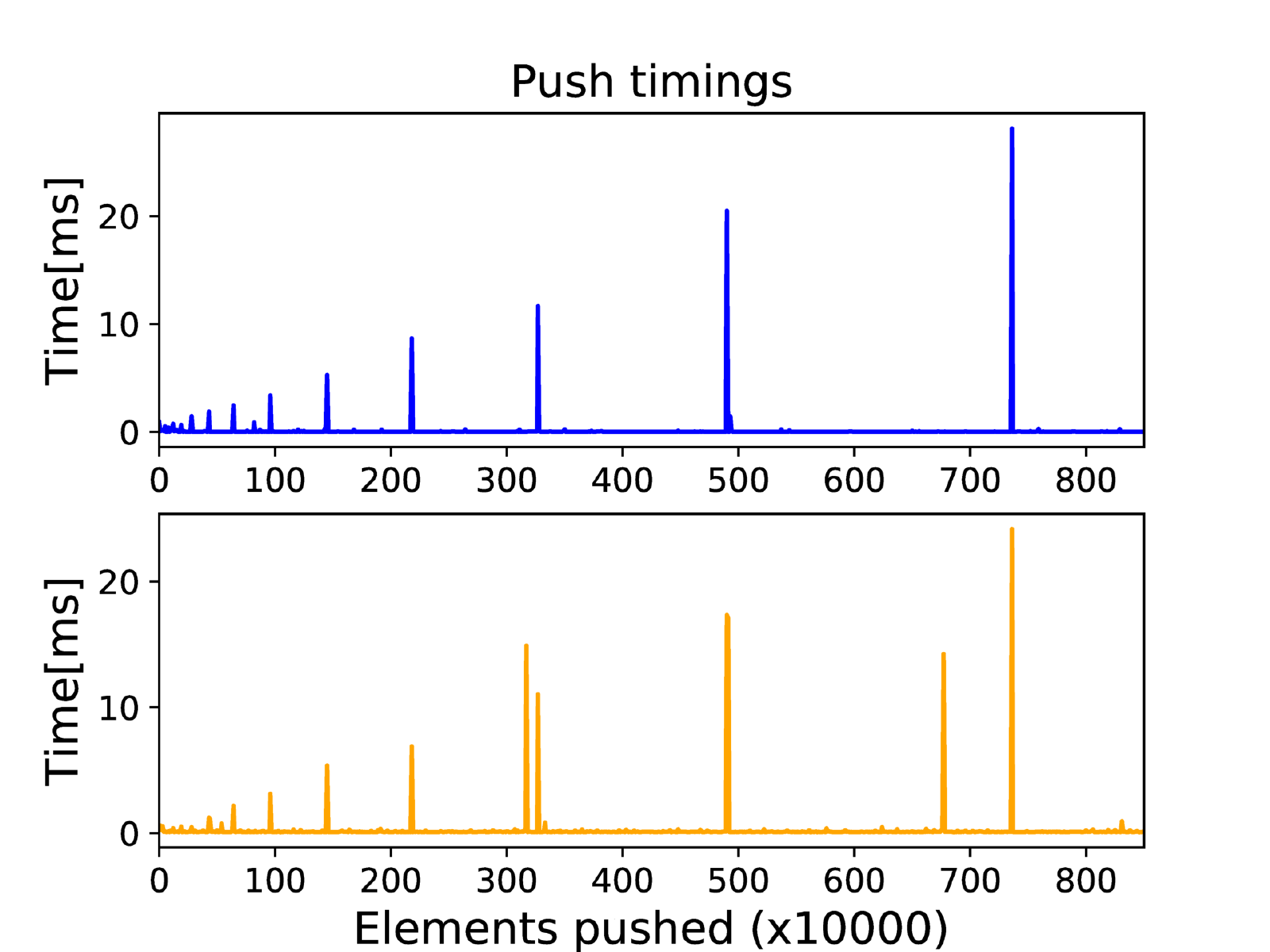}
  \caption{Push timings with native Chrome (top),
    and with \crz at \polhigh level (bottom).}
  \label{fig:push}
\end{figure}

\noindent \cref{fig:push} shows the insertion times for elements, using both a high
resolution timer (top)  and \crz's fuzzy timer (bottom). As can be seen, some
array insertions are slower than others.  We verify that these
additional time costs happened at a point where the buffer allocated by V8 to
support the array \texttt{array} was exhausted, forcing V8 to allocate a new
memory space using using \cref{eqn:v8-resize}.

Observing \cref{fig:push}, the time required to handle the element addition at
the point of buffer exhaustion increases as the size of the array grows. This
is expected as more elements need to be copied by V8 as the buffer grows.
However, as the number of elements added to the array is attacker-controlled,
we can  make \texttt{Array.push} take an arbitrary amount of time.

We exploit this property to mount an attack against \crz's Buffer ASLR policy
despite \crz's attempts at reducing the resolution of \js timers. More
specifically, after a sufficient number of iterations of the loop in \cref{line:loop}, the
time taken to handle the re-allocation of \texttt{array} during the insertion
of an additional element in \cref{line:push} becomes visible despite \crz's low
resolution timer. To deduce the buffer's offset generated by \crz, we apply \crz's buffer ASLR
policy to \cref{eqn:v8-resize} to obtain the following equation.

\begin{equation}
  \label{eqn:v8-resize-aslr}
  \mnewsize + \moffset = (\msize + \moffset) + (\msize + \moffset) \gg 1 + 16.
\end{equation}
Observing the spikes in  \cref{fig:push}, an attacker can detect when the memory of
\texttt{array} is exhausted. From that, to recover the value of $\moffset$, we
rearrange \cref{eqn:v8-resize-aslr} as
\begin{equation}
  \label{eqn:v8-resize-rearranged}
  \moffset = 2 \times \mnewsize - 3 \times \msize - 2 \times 16,
\end{equation}
where $\msize$ and $\mnewsize$ are the size's of \texttt{array} before and
after resizing. Finally, to detect resizing events, an attacker can observe spikes in \cref{fig:push}. Thus, \crz's buffer
ASLR policy can be defeated using two sequential resizing events and applying
\cref{eqn:v8-resize-rearranged} to solve for \textit{offset}.

\begin{table*}[t] {\small
		
		\begin{centering}
			\begin{tabular}{@{}lrcrrrrrcrrrrr@{}}
				\toprule
				& \multicolumn{1}{l}{Temporal}
				&                                & \multicolumn{5}{c}{Top-1 Accuracy (\%)}
				&                                & \multicolumn{5}{c}{Top-5 Accuracy (\%)}                                                                                                                \\
				\cmidrule{4-8}
				\cmidrule{10-14}
				Attack Technique & \multicolumn{1}{l}{Resolution}
				&                                & \poloff                                                & \pollow                                                & \polmed & \polhigh & \polpara
				&                                & \poloff                                                & \pollow                                                & \polmed & \polhigh & \polpara        \\
				
				\midrule
				Cache Occupancy  & 2.9\,ms                        &                                                        & 87.5                                                   & 71.1    & 2.2      & 81.8     & N/A
				&                                & 97.0                                                   & 87.4                                                   & 6.1     & 96.5     & N/A             \\
				Sweep Counting   & 100.0\,ms                      &                                                        & 45.8                                                   & 24.1    & 32.2     & 60.1     & N/A
				&                                & 74.3                                                   & 50.1                                                   & 59.0    & 88.3     & N/A             \\
				DNS Racing       & 20.3\,ms                       &                                                        & 50.8                                                   & 20.9    & 61.1     & 37.2     & 16.2
				&                                & 78.5                                                   & 48.9                                                   & 86.0    & 67.7     & 40.1            \\
				String and Sock  & 1.5\,ms                        &                                                        & 72.0                                                   & 51.3    & 46.2     & 58.4     & 59.9
				&                                & 90.6                                                   & 80.0                                                   & 75.9    & 85.3     & 82.8            \\
				\midrule
				CSS \pp          & 2.8\,ms                        &                                                        & \multicolumn{5}{l}{(with the NoScript extension) 50.1}
				&                                & \multicolumn{5}{l}{(with the NoScript extension) 78.6}                                                                                                 \\
				\bottomrule
			\end{tabular}
		\end{centering}
			\caption{\label{t:fp-results-crz}Closed-world accuracy (percent)
	with different API restriction levels (Intel i5-3470).}
	}
\end{table*}

\subsection{Attacking \crz}
We now present the classification results of the attacks described in \cref{s:cache-attacks-without-timers} across different \crz policies,
starting with the closed-world scenario.
\cref{t:fp-results-crz} summarizes the accuracy of our technique, using the Intel i5-3470 setup outlines in \cref{sec:setup}.
Full results, including further experiments and statistical confidence, are
included in \cref{app:detailed}.

\parhead{Cache Occupancy and Sweep Counting.} As we can see, for the basic cache
occupancy attack, \crz policies have varying impact on the attack accuracy.
\pollow has some impact, but the accuracy is still high. \polmed almost
completely blocks the attack, with the accuracy being slightly more than the
base rate. Surprisingly, \polhigh is less effective than the two lower policy
levels, possibly because of its simpler code design, resulting only in a slight decrease in the accuracy compared to no protection
at all. For the sweep counting attack, we see that the accuracy is lower than
that of the basic cache occupancy channel. However, the \polmed policy no longer
breaks the attack. Furthermore, while lower than that of the cache occupancy
attack, the accuracy is still significantly higher than the base rate. Finally,
because these attacks require Worker threads, which are blocked in
\polpara, they both fail in this policy.

\parhead{DNS Racing.}
The DNS Racing technique achieves a moderate accuracy in the range 20\% to 61\%.
As expected for a technique that requires neither timers nor threads, the attack
also works with \polpara policy.

\parhead{String and Sock.}
The results with the String and Sock tend to be better than DNS Racing. In fact,
the results tend to only be slightly inferior to those of the cache occupancy
attack, despite not requiring timers, arrays, or threads. We further observe
that because the attack uses no protected API, the various \crz policies have
only a marginal effect on attack success.

\parhead{CSS \pp.}
As mentioned in \cref{sec:csspp}, our CSS \pp technique does not require \js and is effective even if the attacker's website is banned from executing any \js code (e.g., due to the NoScript extension~\cite{noscript}). In particular, \crz's focus on \js does not effect our CSS \pp technique, leaving CSS \pp completely unmitigated.

\parhead{Discussion.}
Examining the results in \cref{t:fp-results-crz}, we
see that restricting browser APIs such as threads, timers, and array access can thwart the standard Cache Occupancy and Sweep Counting attacks, and can significantly degrade the effectiveness of the DNS Racing attack.
Nevertheless, the two remaining attacks, String and Sock and CSS \pp, are not affected by this browser-based countermeasure, since they do not use any API which is receiving protection.
While there is some variation in accuracy between the different protection modes for String and Sock, this is likely due to the usability and site loading side-effects related to our fortified version of \crz, and not due to any intrinsic protection offered the API limiting approach.
We thus argue that preventing side channels in today's browsers using API modifications is practically impossible. 
Properly preventing leakage would require a more systematic approach which considers the sources of leakage, and not merely the means for measuring it.

\section{Attacking Hardened Browsers}\label{sec:harden}
Having established the feasibility of mounting cache side channel attacks while only having limited (or no) access to the \js API, in this section we proceed to demonstrate the effectiveness of our techniques on two privacy enhanced browsers: Tor~\cite{TorBrowser} and DeterFox~\cite{CaoCLW17}.

\subsection{Attacking the Tor Browser}

The Tor Browser~\cite{TorBrowser} is a highly-modified version of Firefox, designed to offer a high level of privacy even at the cost of usability and performance.
At a high level, the Tor Browser combines two elements to achieve  a higher level of protection compared to other browsers.
First, it hides the user's browsing habits from network adversaries by using the Tor network as an underlying transport layer. Second,
it provides a highly restrictive browser configuration,
designed to limit or disable convenience features that may have a security impact.
In the context of side channel attacks, the Tor Browser
limits the resolution of the timer API to only 100 milliseconds.

In this section we evaluate our attack techniques from within the Tor Browser
and demonstrate that they are possible even
within this restricted environment.
We note that \citet{ShustermanKHMMOY19} have already demonstrated the Sweep Counting attack in the Tor Browser.
We extends that result, demonstrating that making the environment more restrictive
by disabling \js feature does not guarantee protection.

\parhead{Negative Result: DNS Racing and CSS \pp.}
We begin with a negative result, that the CSS \pp attack we designed is not effective
in the Tor Browser.
The cause is that for security reasons, the Tor Browser does not directly resolve DNS requests.
Instead, it asks a Tor exit relay to resolve the name on its behalf.
This extra redirection step adds a very large delay to DNS requests, on the order of hundreds of milliseconds, as well as a high degree of jitter, well beyond what the attack can handle.
This issue also affects the DNS Racing attack, making it inapplicable.

\parhead{Adapting String and Sock to Tor.}
The String and Sock technique described in \cref{sec:stringandsock} uses a high bandwidth WebSockets connection to offload timing measurements to a remote server.
Unfortunately, due to the high round-trip delay of a Tor connection, the bandwidth available to a WebSockets connection over the Tor transport is significantly lower than a connection made over a regular TCP transport.
Effectively the connection operates in a \textit{stop-and-wait} mode, buffering outgoing packets as long as not all previously transmitted packets are acknowledged.
This buffering removes the timing information that the attack needs.

To avoid buffering, we reduce the communication of our String and Sock attack by sending a probe packet only once every $n$ sweeps over the cache, instead of after every sweep.
We experimentally find that $n=72$ provides the best accuracy.

\begin{figure}[htb]
	\centering	\includegraphics[trim={0cm 0cm 0cm 0cm}, height=0.4\linewidth]{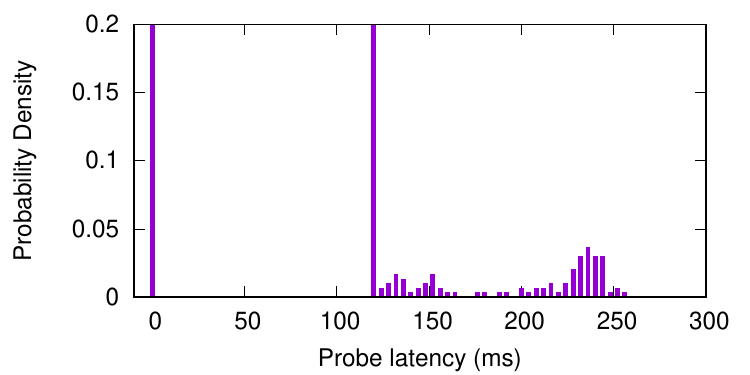}
	\caption{String and Sock Probe latency distribution on Tor Browser using an Intel i5-3470 target (6MB LLC).
		\label{f:tor-sns-x72-histogram}}
\end{figure}

\parhead{Observing the Distribution of Probe Times.}
\cref{f:tor-sns-x72-histogram} shows the probe time distribution using the Intel i5-3470 target. 
As the figure shows, there are three main elements to this distribution.
First, we note a large subset of the probes have a fixed latency of around 120\,ms.
These are buffered by Tor's network layer, as described above, and sent immediately after all previously sent packets are acknowledged.
Thus, these packets do not measure contention of the cache, but instead measure the round-trip delay of the Tor connection.
Next, a large number of probes have a near-zero latency.
These are packets which are sent together with other packets, and similarly do not encode any cache information.
The final subset of the probes has a more diverse set of values, with an estimated mean of between 150 and 250 milliseconds.  These probes encode cache contention information.

\parhead{Website Fingerprinting.}
To demonstrate that these probes indeed contain cache information, we collect a dataset of 10,000 traces of Alexa Top 100 websites on the i5-3470 target running Tor Browser, using our adapted String and Sock method described above. Using this data, we can correctly fingerprint websites, obtaining a   Top-1 accuracy of 20\% and a Top-5 accuracy of 49\%. Well above base rates of 1\% and 5\%, respectively.
This demonstrates that completely eliminating access to timer and array APIs in the Tor Browser does prevent cache attacks.

\subsection{Attacking DeterFox}
DeterFox is a Firefox fork aiming
to provably prevent timing attacks from within browser executed code~\cite{CaoCLW17}.
Its authors argue that when using DeterFox, ``an observer in a JavaScript reference frame will always obtain the same fixed timing information, so that timing attacks are prevented''.
To achieve this, DeterFox splits its execution context into multiple deterministic reference frames, and uses a priority-based event queue for communication between these reference.

However, we note that our CSS \pp technique does not require any \js, with the colluding DNS server providing time measurement remotely.
Thus, our techniques effectively sidestep all of the side channel protections offered by DeterFox.
To  demonstrate the effectiveness of our attacks on DeterFox, we collect one more dataset of 10,000 traces of Alexa Top 100 websites, using the CSS \pp method while using DeterFox. As expected, DeterFox's provably secure deterministic timing countermeasure did not prevent our attack, giving us a Top-1 accuracy of 66\% and a Top-5 accuracy of 88\%.

\section{Conclusion}
This paper shows that defending against \js-based side-channel attacks is more
difficult than previously considered.
We show that advanced variants of the cache contention attack allow \pp attacks
to be mounted through the browser in extremely constrained situations. Cache
attacks cannot be prevented by reduced timer resolution, by the abolition of
timers, threads, or arrays, or even by completely disabling scripting support.
This implies that any secret-bearing process which shares cache resources with a
browser connecting to untrusted websites is potentially at risk of exposure.

We also show that the reduced requirements of our attack make it agnostic across a variety of microarchitectures with no modifications. This allows us to present the first end-to-end side-channel attack which targets Apple's new M1 processors.

So, how can security-conscious users access the web? One complicating factor to
this concept is the fact that the web browser makes use of additional shared
resources beyond the cache, such as the operating system's DNS resolver, the
GPU and the network interface.
Cache partitioning seems a promising approach, either using spatial isolation based on cache coloring~\cite{KimPM12}, or by OS-based temporal isolation~\cite{GeYCH19}.

\ifAnon
\else
  \section*{Acknowledgements}

This work was supported
the Air Force Office of Scientific Research (AFOSR) under award number FA9550-20-1-0425;
an ARC Discovery Early Career Researcher Award (project number DE200101577);
an ARC Discovery Project (project number DP210102670);
the Defense Advanced Research Projects Agency (DARPA) and Air Force Research Laboratory (AFRL) under contracts FA8750-19-C-0531 and HR001120C0087;
Israel Science Foundation grants 702/16 and 703/16;
the National Science Foundation under grant CNS-1954712;
the Research Center for Cyber Security at Tel-Aviv University established by the State of Israel, the Prime Minister's Office and Tel-Aviv University;
and gifts from Intel and AMD.

The authors thank Jamil Shusterman for his assistance in bringing up the measurement setup.

\bibliographystyle{plainnat}

\appendix

\section{Machine Learning Model}\label{app:model}
Our machine learning classifier receives as input a side-channel trace, and
outputs a probability distribution over the 100 potential websites. Before the
trace is fed to the model, the input vector was normalized between 0 and 1. We
then used a deep learning network to perform our analysis, meaning that feature
extraction was done inside the neural network and did not require additional
preprocessing steps. We used the deep learning model whose hyperparameters are
presented in \cref{tab:hyperparameters}. The model begins with a convolution
layer which learns the unique patterns of each label, followed by a Max-Pooling
layer which reduces the dimensionality of the output of the previous layer. The
output of the Max-Pooling layer is then reshaped to a one dimension vector and
fed to a Long-Short Term Layer, which extracts temporal features over its input.
Finally, the output layer of the network is a fully-connected layer with a
softmax activation function.

\begin{table}[htb] {\footnotesize
    \centering
    \caption{Hyperparameters for the deep learning classifier \label{tab:hyperparameters}}
    \begin{tabular}{ll}
      \toprule
      \textbf{Hyperparameter} & \textbf{Value}                      \\
      \midrule
      Optimizer               & Adam                                \\
      Learning rate           & 0.001                               \\
      Batch size              & 128                                 \\
      Training epoch          & Early stop by validation accuracy   \\
      Input units             & vector size of the 30 seconds input \\
      Convolution layers      & 2                                   \\
      Convolution activation  & relu                                \\
      Convolution Kernels     & 256                                 \\
      Convolution Kernel size & 16,8                                \\
      Pool size               & 4                                   \\
      LSTM activation         & tanh                                \\
      LSTM units              & 32                                  \\
      Dropout                 & 0.7                                 \\
      \bottomrule
    \end{tabular}
  }
\end{table}

The model was evaluated on a test set whose traces are not part of the training
set. The metric we use is accuracy  -- the probability of a trace to be
classified correctly. To avoid overfitting in model estimation, we employ 10
fold cross validation, a method which divides the dataset into 10 parts, with
each part becoming the test set while the others are used as the train set. Each
training set is fed to a different model, and the evaluation is made on the
related test set. After each experiment, we noted the average cross-fold
accuracy, as well as the standard deviation between folds.

The output of our classifier is not only the label of the most probable class,
but rather a complete probability distribution over all possible labels. This
flexibility allows us to capture the case where the attacker has some prior
knowledge of the victim and some expectation of the websites they may be
browsing. To do so, we look not only at the top-rated label, but also at a few
of the next most probable predictions.  This methodology was previously used in
similar works where low-accuracy classifiers were
evaluated~\cite{Caliskan-IslamHLNVYG15,NarayananPGBSSS12}. We thus calculated
not only the raw accuracy, but also the probability that the right prediction is
among the top 5 websites output as the most probable by the classifier. The base
accuracy rate of this prediction method, as obtained by a random classifier with
no knowledge of the traces, is 5\%.

The machine learning model was implemented in python version 3.6, using
TensorFlow~\cite{tensorflow2015-whitepaper} library version 1.4. The model
training algorithms were run on a cluster made out of Nvidia GTX1080 and GTX2080
graphics processing units (GPUs), managed by Slurm workload manager~\cite{slurm}
version 19.05.4.

\section{Detailed Results}\label{app:detailed}
\cref{t:fingerprinting-results} and \cref{t:fingerprinting-top5} list the accuracies
of our website fingerprinting attacks using different techniques, including the
standard deviations calculated over the 10 folds. We also list the performance
of the String and Sock attack in case the remote server is located on a
different continent.

\cref{t:benchmark_latency} and \cref{t:benchmark_throughput} list individual
benchmark results for JetStream~1.1's Latency and Throughput benchmarks. We
describe our modifications to \crz in \cref{sec:crzevaluation} and compare it to
the baseline (No extensions running).

\begin{table}[htb]{
    \begin{centering}
      \caption{\crz JetStream~1.1 Latency Benchmarks.}
      \label{t:benchmark_latency}
      \begin{tabular}{@{}lll@{}}
        \toprule
                          & {Modified}     & {No Extensions} \\
        \midrule
        3d-cube           & 40.70 ± 3.400  & 12.64 ± 0.5798  \\
        base64            & 13.78 ± 1.262  & 14.26 ± 0.7038  \\
        cdjs              & 103.4 ± 16.04  & 102.6 ± 9.810   \\
        code-first-load   & 115.9 ± 1.429  & 410.9 ± 5.332   \\
        code-multi-load   & 107.1 ± 2.784  & 386.6 ± 5.677   \\
        crypto-aes        & 40.45 ± 1.248  & 3.584 ± 0.0846  \\
        crypto-md5        & 19.72 ± 0.4929 & 8.249 ± 0.0708  \\
        crypto-sha1       & 10.46 ± 0.1325 & 3.798 ± 0.0629  \\
        date-format-tofte & 37.00 ± 1.537  & 38.29 ± 0.2797  \\
        date-format-xparb & 45.24 ± 3.821  & 45.86 ± 1.948   \\
        n-body            & 41.76 ± 0.4176 & 13.89 ± 0.6885  \\
        regex-dna         & 130.5 ± 1.550  & 129.8 ± 2.662   \\
        splay-latency     & 1199 ± 57.44   & 1007 ± 31.18    \\
        tagcloud          & 57.86 ± 2.163  & 57.18 ± 2.100   \\
        typescript        & 70.78 ± 2.753  & 41.95 ± 1.294   \\

        \bottomrule
      \end{tabular} \\
    \end{centering}
  }
\end{table}

\begin{table}[htb]{
    \begin{centering}
      \caption{\crz JetStream~1.1 Throughput Benchmarks.}
      \label{t:benchmark_throughput}
      \begin{tabular}{@{}lll@{}}
        \toprule
                        & {Modified}     & {No Extensions} \\
        \midrule
        bigfib.cpp      & 357.7 ± 30.09  & 1.087 ± 0.0035  \\
        box2d           & 159.7 ± 7.584  & 21.37 ± 0.5126  \\
        container.cpp   & 314.9 ± 2.921  & 2.797 ± 0.0220  \\
        crypto          & 151.7 ± 1.498  & 0.6892 ± 0.0008 \\
        delta-blue      & 250.3 ± 7.558  & 3.042 ± 0.0277  \\
        dry.c           & 186.8 ± 2.808  & 0.2445 ± 0.0044 \\
        earley-boyer    & 118.9 ± 0.6421 & 33.76 ± 1.295   \\
        float-mm.c      & 368.9 ± 16.68  & 2.333 ± 0.0381  \\
        gcc-loops.cpp   & 408.2 ± 137.1  & 0.8519 ± 0.0157 \\
        hash-map        & 197.0 ± 16.16  & 10.69 ± 0.2144  \\
        n-body.c        & 200.9 ± 4.306  & 0.6504 ± 0.0067 \\
        navier-stokes   & 212.2 ± 3.232  & 0.6890 ± 0.0069 \\
        proto-raytracer & 188.9 ± 4.286  & 44.10 ± 1.225   \\
        quicksort.c     & 264.2 ± 8.418  & 3.272 ± 0.0198  \\
        regexp-2010     & 321.1 ± 6.017  & 322.8 ± 2.092   \\
        richards        & 153.1 ± 2.395  & 16.13 ± 0.6450  \\
        splay           & 248.1 ± 15.26  & 254.8 ± 9.464   \\
        towers.c        & 235.0 ± 3.063  & 0.9661 ± 0.0396 \\
        zlib            & 267.8 ± 3.097  & 1.083 ± 0.0079  \\
        \bottomrule
      \end{tabular}\\
    \end{centering}
  }
\end{table}

\end{document}